\documentclass[a4paper,11pt]{article}
\usepackage{jheppub} 
\usepackage{lineno}
\usepackage[caption=false]{subfig}

\usepackage{mathrsfs}
\usepackage{physics}
\usepackage{tikz}
\usepackage[caption=false]{subfig}
\usepackage[colorlinks=true,linkcolor=blue,citecolor=blue,urlcolor=blue]{hyperref}
\usepackage{cleveref}
\usepackage{amsmath,amssymb,amsfonts,amsthm}
\usepackage{graphicx}
\usepackage{ragged2e}
\usepackage{float}
\usepackage{bm}
\usepackage{svg}
\usepackage{extarrows}

\usepackage{dsfont}
\usepackage[title]{appendix}
\newcommand{\dbar}{d\hspace*{-0.08em}\bar{}\hspace*{0.1em}}
\usepackage{hyperref}

\def\rar{\rightarrow}

\def\bea{\begin{eqnarray}}
\def\eea{\end{eqnarray}}
\def\be{\begin{equation}}
\def\ee{\end{equation}}

\usepackage{filecontents}

\makeatletter
\renewcommand{\note}[1]{%
    \begingroup
    \renewcommand{\thefootnote}{}
    \footnote{#1}%
    \endgroup
}
\makeatother

\title{\boldmath Quantum Mpemba-like effect in Unruh thermalization}

\author[a,\dagger]{Zihao Wang\note{$^\dagger$ These authors contributed equally to this work.},}
\author[a,\dagger]{Wenjing Chen,}
\author[a,\dagger]{Si-Wei Han,}
\author[a]{Xiaoshan Feng,}
\author[a,c]{Linmu Qiao,}
\author[d]{Zhichun Ouyang,}
\author[a,b,1]{and Jun Feng\note{$^1$ Corresponding author.}}
\affiliation[a]{School of Physics, Xian Jiaotong University, Xi'an 710049, Shaanxi, China}
\affiliation[b]{Hefei National Laboratory, Hefei 230088, China}
\affiliation[c]{Department of Mechanical and Automation Engineering,
The Chinese University of Hong Kong, Shatin, Hong Kong, China}
\affiliation[d]{Department of Physics, Hong Kong University of Science and Technology,
Clear Water Bay, Hong Kong, China}

\emailAdd{j.feng@xjtu.edu.cn}

\abstract{We revisit the thermal nature of the Unruh effect within a quantum thermodynamic framework. For a Unruh-deWitt (UDW) detector in $n$-dimensional Minkowski spacetime, we demonstrate that its irreversible thermalization to a Gibbs equilibrium state follows distinct trajectories on the Bloch sphere, which depend on the types of fields the detector interacts with, as well as the spacetime dimensionality. Using thermodynamic process functions, particularly quantum coherence and heat that form the quantum First Law, we characterize the Unruh thermalization through a complementary time evolution between the trajectory-dependent rates of process functions. Grounded in information geometry, we further explore the kinematics of the detector state as it “flows” along the trajectory. In particular, we propose two heating/cooling protocols for the UDW detector undergoing Unruh thermalization. We observe a quantum Mpemba-like effect, characterized by faster heating than cooling in terms of Uhlmann fidelity "distance" change. Most significantly, we establish the maximum fidelity difference as a novel diagnostic that essentially distinguishes between Unruh thermalization and its classical counterpart, i.e., classical thermal bath-driven thermalization of an inertial UDW detector. This compelling criterion may serve as a hallmark of the quantum origin of the Unruh effect in future experimental detection and quantum simulation. Finally, we conclude with a general analysis of Unruh thermalization, starting from equal-fidelity non-thermal states, and demonstrate that the detectors' fidelity and "speed" of quantum evolution still exhibit a Mpemba-like behavior.

 }

\begin{document}
\maketitle
\flushbottom
\section{Introduction}
The Unruh effect predicts that a detector undergoing uniform proper acceleration ($a = const$) in Minkowski spacetime will perceive the quantum vacuum as a thermal state with temperature $T_U=a/2\pi$ \cite{sec1-1}. This phenomenon demonstrates the observer-dependent nature of particle content in quantum field theory (QFT), as noted by Fulling \cite{sec1-2-1} and Davies \cite{sec1-2-2}. Furthermore, from the perspective of the Equivalence Principle, the Unruh effect provides a simplified yet tractable platform for exploring other quantum gravity phenomena, such as the Hawking effect of black holes \cite{sec1-3} or cosmological radiation \cite{sec1-4}. 

Despite its foundational importance, the direct experimental verification of the Unruh effect remains a formidable challenge, lying at the cutting edge of current technology. This is because even creating an exceedingly small Unruh temperature requires huge acceleration (e.g., reaching a temperature $T_U\sim 1 \mathrm{K}$ corresponds to a linear acceleration of order $10^{21} \mathrm{~m} / \mathrm{s}^2$). Although numerous experimental proposals have been put forward \cite{v2-1,v2-2,v2-3,v2-4,v2-5}, conclusive evidence of the Unurh effect remains elusive. In this context, analogue quantum simulators have emerged as a promising alternative, offering a flexible and well-controlled laboratory setting for studying relativistic QFT effects \cite{v2-6}. In these systems, a characteristic speed (such as sound speed \cite{v2-7} or matter-wave speed \cite{v2-8}) replaces the speed of light as the propagation speed for an effective massless field. Consequently, the dynamics of perturbations are expected to mimic phenomena like Hawking radiation \cite{v2-9} or the Gibbons-Hawking effect \cite{v2-10}. Recently, several quantum simulations of the Unruh effect have been reported in Bose-Einstein condensate (BEC) systems \cite{v2-11,v2-12}.

The primary goal of most experimental simulations is to verify the \emph{thermal nature} of the Unruh effect, operationally defined by the attainment of a Planckian response spectrum upon the detector’s equilibration. Moreover, authenticating a \emph{genuine} Unruh effect requires additional verification of unambiguous signatures that distinguish the quantum-originated Unruh radiation from classical thermal radiation at temperature $T_U$. For instance, in BEC simulation \cite{v2-11}, long-range phase coherence and temporal reversal of the matter-wave radiation imitate the spatial and temporal coherence of Unruh radiation. However, two main obstacles exist \cite{sec1-7-5}: (1) no simulation of exotic scenarios exhibiting non-Planckian responses (e.g., anyonic response spectrum associated with a conformal background \cite{sec1-7-6}) at the detector's equilibrium has been reported, and (2) the existence of a universal and compelling quantum-discriminating signature for the Unruh effect is still debated.

Theoretically, the Unruh effect is derived from the thermalization theorem \cite{sec1-6}, which guarantees that the detector satisfies the Kubo-Martin-Schwinger (KMS) condition, a hallmark of thermal equilibrium \cite{sec1-7}. Crucially, for the Unruh effect, this does not preclude deviations from strict Planckian statistics in a detector's response. A striking manifestation of this is the phenomenon of \emph{statistics inversion} \cite{sec1-7-1}, exhibited by an Unruh-deWitt (UDW) detector \cite{sec1-5} coupled to a massless scalar field in $n$-dimensional spacetime. Specifically, the detector's excitation spectrum switches from Bose-Einstein statistics for even $n$ to Fermi-Dirac statistics for odd $n$. Historically, this inversion challenged a thermal interpretation of the Unruh effect \cite{sec1-7-3}. However, within the framework of the thermalization theorem, the statistics inversion is fully consistent with the KMS condition and thus reinforces, rather than invalidates, this interpretation \cite{sec1-7-4,sec1-7-7}. Nevertheless, it sharpens the distinction between the Unruh radiation and a classical thermal bath at the same temperature.

Before reaching equilibrium with the background fields, the detector undergoes an irreversible (Unruh) thermalization process. Its final response spectrum thus reflects only asymptotic equilibrium properties, while all information about the intermediate quantum dynamics is erased. For example, even when the (Unruh) temperature uniquely determines the final thermalized state, the open quantum dynamics of a UDW detector \cite{sec1-8} allows itself to evolve from an arbitrary initial state along distinct thermalization trajectories toward this equilibrium. Quantified by trajectory-dependent \emph{process functions} (e.g., quantum Fisher information), it was shown \cite{sec1-7-8} that statistics inversion does not provoke any dependence of thermalization trajectory on the parity of spacetime dimension. This finding is consistent with the established thermal interpretation of the Unruh effect.

The ensemble of possible thermalization trajectories is governed by the open quantum dynamics of the detector, which in turn encodes details of the coupled background field. This insight provides a framework for distinguishing Unruh thermalization from a thermal bath-driven process by comparing these trajectory-dependent process functions in both scenarios. Early studies \cite{sec1-7-10,sec1-7-9} examined subtle asymmetries in specific process functions (such as quantum coherence monotones, geometric phase, and entanglement measures) between the Unruh and classical thermal baths. However, numerical differences alone are unsatisfactory to regard these quantities as compelling and unambiguous signatures of the quantum-originated Unruh effect.

In this paper, we revise the thermalization process of an accelerating UDW detector in $n$-dimensional Minkowski spacetime. Since the detector undergoes an irreversible process driven by a quantum (Unruh) effect, we argue that a reexamination of its non-equilibrium dynamics through the lens of modern quantum thermodynamics \cite{sec1-9,sec1-10,sec1-10+} is both legitimate and timely. We expect this synthesis to spark new insights into the thermal nature of the Unruh effect. Specifically, we address the two key issues: (1) characterizing the Unruh thermalization process via trajectory-dependent \emph{thermodynamic process functions} that articulate thermodynamic laws in the quantum regime \cite{sec1-11,sec1-12}, and (2) presenting novel diagnostic signatures of the Unruh-effect-induced thermalization that distinctly set it apart from a classical bath-driven thermalization at the same temperature.

The paper is structured as follows. In Section \ref{sec2}, we review the open quantum dynamics of an accelerating UDW detector in $n$-dimensional Minkowski spacetime. We assume a weak coupling to typical background fields (e.g., a massless scalar field, or a classical thermal field), which allows the detector’s density matrix to be described by a Lindblad-form master equation that evolves irreversibly. In particular, for a massless background, the statistics inversion encoded in the detector response function is discussed. 

In Section \ref{sec3}, we show that the irreversible thermalization of a UDW detector is characterized by a one-way trajectory in the Bloch sphere, which depends on the types of field the detector interacts with, as well as the spacetime dimensionality. We are particularly interested in two types of process: \emph{Unruh thermalization} of an accelerating detector interacting with a massless scalar field, as well as the thermalization process undergone by an inertial detector driven by a classical thermal bath. By employing the change rates of quantum coherence and heat, which are used to formulate a quantum version of the First Law, \cite{sec1-11}, we confirm that bath-driven thermalization has a significantly longer timescale than Unruh thermalization. We introduce the difference between the change rates of quantum coherence and heat as a diagnostic signature, whose maximum value becomes larger for higher-dimensional spacetimes in Unruh thermalization, while it decreases for bath-driven thermalization. This is our first key result.

Another central result is presented in Section \ref{sec4}. We design a heating/cooling protocol for a UDW detector (see Fig.\ref{Protocol}(a)) initialized in a Gibbs state, which then undergoes Unruh thermalization toward equilibrium at fixed Unruh temperature. From the perspective of quantum thermal kinematics \cite{sec1-15}, we demonstrate a notable asymmetric propagation of the state point along the trajectory in the Unruh thermalization process (Section \ref{4.2.1}): the trajectory to thermal equilibrium is inherently different depending on whether the detector is heating up or cooling down. We uncover a \emph{quantum Mpemba-like effect} (QME) \cite{Mpemba0,sec1-16,Mpemba1} in Unruh thermalization wherein the detector always heats up faster than it cools down, for all types of background fields. Most importantly, we find that the maximum difference in state distance, measured by the Uhlmann fidelity, between the heating and cooling processes serves as a compelling diagnostic signature to distinguish Unruh thermalization from a thermal radiation-driven thermalization. We observe an unexpected dependence of this maximal fidelity difference on the parity of the spacetime dimension for thermalization driven by a classical field, which is entirely absent in quantum-originated Unruh thermalization. In Sections \ref{4.2.2} and \ref{sec4.3}, we further identify a similar asymmetry of Unruh thermalization for a three-temperature protocol involving two UDW detectors (see Fig.\ref{Protocol}(b)), as well as for detectors starting from non-thermal initial states.

In Section \ref{sec5}, we summarize our results and discuss their potential applications, for example, in future quantum simulations of the Unruh effect.

Throughout this work, we adopt natural units: $ h = c = k _ { B } = 1$

\section{Open dynamics in $n$-dimensional flat spacetime} 
\label{sec2}

We begin by reviewing the open quantum dynamics of a uniformly accelerating UDW detector in $n$-dimensional Minkowski spacetime, following the framework established in \cite{sec1-7-8}. The detector, modeled as a two-level atom, is weakly coupled to a bath of fluctuating quantum scalar fields. Its reduced dynamics are governed by a Lindblad-form master equation, which describes irreversible evolution toward thermal equilibrium. 

The total Hamiltonian of the detector-field system is:
\begin{equation}
    {H}_{\text{tot}}={H}_{\text{d}}+{H}_{\Phi}+\mu {H}_{\text{int}},
\end{equation}
where the two-level atom Hamiltonian is ${H}_{\text{d}}=\frac{1}{2} \omega \sigma_3$, and $H_{\Phi}$ is the free Hamiltonian of a background scalar field $\Phi(x)$. For simplicity, we assume the UDW detector is pointlike and couples linearly to the scalar field via the interaction Hamiltonian ${H}_{\text{int}}=(\sigma_+ +\sigma_-)\otimes\Phi(x)$, where $x=x(t)$ denotes the time-dependent trajectory of the detector\footnote{For finite-size detector, a spatial smearing function $F(\bm y)$ can be incorporated by replacing $\Phi(x)$ with a smeared operator, e.g.\ $\int d^{3}\bm y\,F(\bm y)\,\Phi\!\bigl(x,\bm y\bigr)$ in an appropriate frame (see \cite{rev-4,rev-5} for example).}. Standard quantum mechanics ensures that the detector-field composed Hamiltonian evolves following $\dot{\rho}_{\text{tot}}( t )=-i[{H}_{\text{tot}},\rho_{\text{tot}}( t )]$, the von Neumann equation of total density matrix.

In the weak-coupling regime ($\mu\ll 1$),  the state of the composed system can be Born-approximated as $\rho_{tot}(t)\approx \rho_{\text{d}}( t )\otimes\rho_{\Phi}$. With in mind that the typical evolution timescale of the detector now in the interaction picture is of order $\mathcal{O}(1/\mu^2)$. Assuming that the field correlation function decays sufficiently rapidly, the memory effect in the detector's exact dynamics can then be neglected via Markov approximation \cite{sec1-8}. This allows one to trace out the field degrees of freedom and obtain the Gorini-Kossakowski-Sudarshan-Lindblad (GKSL) master equation \cite{sec2-1,sec2-2} for the detector’s density matrix $\rho$ as:
\begin{equation}
\frac{d\rho}{dt} = -i[H_{\text{d}} + H_{\text{LS}}, \rho] + \mathcal{L}[\rho],
\label{master}
\end{equation}
where $H_{\text{LS}}$ is the Lamb shift Hamiltonian accounting for environment-induced energy renormalization, and $\mathcal{L}[\rho]$ is the dissipator that encodes the irreversible dynamics induced by the environment. 

As proven in Davies' seminal works \cite{rev-2,rev-3}, the master equation for open dynamics \eqref{master} becomes accurate in the van Hove limit \cite{rev-1} [i.e., $\mu^2 t \sim \mathcal{O}(1)$ while taking $\mu \rightarrow 0$ and $t \rightarrow \infty$ simultaneously], as all higher-order terms in the perturbation series of evolution tend to zero then. As a result, this approach does not inherently restrict other parameters, notably the detector's characteristic frequency $\omega$ or the Unruh temperature $T$. Even when relaxed to a more physical limit, i.e., $g^2 t \sim \mathcal{O}(1)$ with small but nonvanishing $g$, the validity of the quantum master could be guaranteed by some reliability conditions of Markov approximation \cite{sec2-plus3} that are well-justified in the sufficiently weak-coupling regime\footnote{It is fair to mention that recent studies \cite{sec2-plus1,sec2-plus2} have proposed an alternative Markov approximation, resulting in a quantum master equation different from \eqref{master} that we used. Nevertheless, the validity of that alternative approach is generally restricted to high-temperature regimes. Since we do not wish to impose prior restrictions on the Unruh temperature, we adhere to the conventional Markov approximation based on the van Hove limit. Within this framework \cite{sec2-plus3}, the weak-coupling limit alone is sufficient to ensure the validity of the master equation \eqref{master}.}.

For the two-level UDW detector, the dissipator can be written explicitly as
\begin{equation}\label{Lindblad operator}
      \mathcal{L}\left[ \rho \right] = \sum _ { i , j = 1 } ^ { 3 } C _ { i j } \left[ \sigma _ { j } \rho \sigma _ { i } - \frac { 1 } { 2 } \left\{ \sigma _ { i } \sigma _ { j } , \rho \right\} \right]
\end{equation}
where the Kossakowski matrix $C_{ij}$ takes the form:
\begin{equation}
     C _ { i j } = \frac { \gamma _ { + } } { 2 } \delta _ { i j } - i \frac { \gamma _ { - } } { 2 } \epsilon _ { i j k } n _ { k } + \gamma _ { 0 } \delta _ { 3 , i } \delta _ { 3 , j }
\end{equation}
with coefficients
\begin{equation}
     \gamma _ { \pm } := \mu^2\left[\mathcal{G} ( \omega ) \pm  \mathcal{G} ( - \omega )\right] ,~~~~~ \gamma _ { 0 } := \mu^2\mathcal{G} ( 0 ) - \gamma _ { + } / 2
     \label{gamma}
\end{equation}
defined in terms of the response function $\mathcal{G}$, which is the Fourier transform of Wightman function evaluated along the detector’s trajectory:
\be
 \mathcal{G}(\omega)= \int_{-\infty}^{\infty} \mathrm{d}s\, e^{ i \omega s}\langle\Phi(x(s)) \Phi(x(0))\rangle
 \label{response}
\ee 

Using the Bloch representation of the density matrix
\begin{equation}
     \rho (  t  ) = \frac { 1 } { 2 } \left( 1 + \sum _ { i = 1 } ^ { 3 } n_ { i } (  t  ) \sigma _ { i } \right),
     \label{bloch}
\end{equation}
the GKSL master equation reduces to a dynamical equation for the Bloch vector $\bm{n}=\left(n_1,n_2,n_3\right)^{\intercal}$, which is a Schr\"odinger-like equation $ \dot{\boldsymbol r} + 2 \boldsymbol{M} \cdot \boldsymbol{r} + \boldsymbol{\eta} = 0$, where\footnote{For more general model with detector Hamiltonian ${H}_{\text{d}}=\omega \boldsymbol{m}\cdot\bm{\sigma}/2$, the dissipator can be explicitly calculated as \cite{sec1-14} $\mathcal{L}[\rho]=\left[\gamma_0(\boldsymbol{m} \cdot \boldsymbol{n})-\gamma_{-}\right] \boldsymbol{m} \cdot \boldsymbol{\sigma}-\left(\gamma_{+}+\gamma_0\right) \boldsymbol{n} \cdot \boldsymbol{\sigma}$. Specific case $\boldsymbol{m}=(0,0,1)$ leads to the vector equation of $\bm{n}$ and $\bm{M}$.} 
%
%
%
%
%
$\boldsymbol{\eta}=(0,0,-2\gamma_{-})$ and 
\begin{equation}
    \boldsymbol{M} := 
    \left(\begin{array}{ccc}
     \gamma _ { + } + \gamma _ { 0 } & {\Omega} / 2 & 0 \\ - {\Omega} / 2 & \gamma _ { + } + \gamma _ { 0 } & 0 \\ 0 & 0 & \gamma _ { + }
     \end{array}\right).
     \label{M}
\end{equation}
Here, $ { \Omega } = \omega + i \left[ \mathcal{K} ( - \omega ) - \mathcal{K} ( \omega ) \right] $ is the Lamb-shifted energy gap, with $\mathcal{K} ( \lambda ) = \frac { 1 } { i \pi } P \int _ { - \infty } ^ { \infty } d \omega \frac { \mathcal{G}  ( \omega ) } { \omega - \lambda }$ being the Hilbert transform of $\mathcal{G}(\omega)$. In practice, the Lamb shift is often negligible compared to the bare energy gap.

Starting from a state with vector length $l_0$ (for mixed state $l_0<1$, for pure state $l_0=1$) and angle $\Theta_0$ to the $z$-axis, the general solution for the Bloch vector components of the density matrix \eqref{bloch} can be explicitly given as:
 \begin{equation}\label{solution}
\left\{\begin{aligned}
n _ { 1 } (  t  ) &= \mathcal{E}_1( t ) l_0 \sin \Theta_0 \cos \Omega t \\ 
n _ { 2 } (  t  ) &= \mathcal{E}_1( t ) l_0 \sin \Theta_0 \sin \Omega t \\ 
n _ { 3 } (  t  ) &=  \mathcal{E}_2( t )\left(l_0 \cos \Theta_0+\gamma\right)-\gamma
     \end{aligned}
     \right.
 \end{equation}
 where the decay rate parameters $\mathcal{E}_1( t ) := \exp \left[-2 \left(\gamma_{+}+\gamma_0\right)  t \right]$ and $\mathcal{E}_2( t ) := \exp \left(-2  \gamma_{+}  t \right)$, as well as a ratio $\gamma \equiv \gamma_{-} / \gamma_{+}$ have been introduced. The length of the Bloch vector evolves as:
 \be
 l( t ):=\sqrt{\sum_{i=1}^3n_i^2}=\sqrt{\left[\mathcal{E}_2\left(l_0 \cos \Theta_0+\gamma\right)-\gamma\right]^2+\mathcal{E}_1^2 l_0^2 \sin ^2 \Theta_0} 
 \label{length}
 \ee
 
 It is not hard to see that after a sufficiently long time $ t \rar\infty$, the UDW detector is thermalized to a unique end, i.e., a Gibbs state
  \begin{equation}\label{rho_equil}
      \rho_ {\text{eq}} ( \beta ) = \frac { 1 } { 2 }
      \left(\begin{array}{cc}
      1 - \gamma & 0 \\ 0 & 1 + \gamma
      \end{array}\right)
        =\frac{e^{-H_{\text{d}}/T_{\text{eff}}}}{\text{Tr}\left[e^{-H_{\text{d}}/T_{\text{eff}}}\right]}
 \end{equation}
at an effective temperature
\be
T_{\text{eff}}=\frac{\Omega}{2 \tanh ^{-1}(\gamma)}
\ee 
which is irrelevant to the detector's initial state but solely determined by the KMS condition of environment fields \cite{sec1-7-7}, i.e., $\mathcal{G}(-\omega)=e^{-\omega/T_{\text{eff}}} \mathcal{G}(\omega)$. However, the detector can follow different paths within its Hilbert space to reach the unique equilibrium, which are expected to encode the dynamic features of the detector and the background field, as well as the space-time geometry, including its dimensionality. 

    

We now specialize to a UDW detector on the uniformly accelerated trajectory:
\begin{equation}
     x ^ { 0 }= a ^ { - 1 } \sinh  a  t  , ~~x ^ { 1 }= a ^ { - 1 } \cosh a  t , ~~
     x^i= 0~~~~(i=2,\cdots,n-1)
     \label{trajectory}
\end{equation}
where $ t $ is detector proper time and $a$ denotes the acceleration. Along the trajectory, the scalar field interacting with the detector induces a response function in general, like:
\begin{equation}\label{G_omega massless}
     \mathcal{G} _ { n } ( \omega ) =  \frac { \pi } { \omega } \frac { D _ { n } ( \omega ) } { 1-e ^ { - \beta \omega }  }
\end{equation}
where the profile function $D_n(\omega)$ is an integral of the modified Bessel function:
\begin{equation}
     D _ { n } ( \omega ) = \frac { 2 } { \pi | \Gamma \left( \frac{i \omega} {a} \right) | ^ { 2 } } \int \frac { d ^ { n - 2 } k } { ( 2 \pi ) ^ { n - 2 } } \left| K _ { \frac{i \omega} {a} } \left( \frac{\sqrt { m ^ { 2 } + | \boldsymbol{k} | ^ { 2 } }} {a} \right) \right| ^ { 2 }
\end{equation}
and the Unruh temperature is $\beta:=1/T_U=a/2\pi$.

For a massless field ($m=0$), the response function \eqref{response} simplifies to:
\begin{equation}\label{massless response fun}
    \mathcal{G} _ { n } ^ { m = 0 } ( \omega ) = \frac { \pi ^ { \frac { n - 5 } { 2 } } \beta ^ { 3 - n } } { 2 \Gamma ( \frac { n - 1 } { 2 } ) } \left| \Gamma ( \frac { n } { 2 } - 1 + \frac { \beta \omega } { 2 \pi } i ) \right| ^ { 2 } \frac { f _ { n } ( \omega ) } { e ^ { - \beta \omega } - ( - 1 ) ^ { n } }
\end{equation}
where 
\begin{equation}
     f _ { n } ( \omega ) = \left\{\begin{matrix} - \sinh ( \beta \omega / 2 ) \quad $if n is even$  \\ \cosh ( \beta \omega / 2 ) \quad $if n is odd$ \end{matrix} \right.
\end{equation}
This exhibits the exotic “statistics inversion”: the detector responds with Bose–Einstein statistics for even $n$ and Fermi–Dirac statistics for odd $n$, despite the bosonic nature of the field. Although lacking a Planck factor, the KMS condition $\mathcal{G}(-\omega)=e^{-\omega/T_{\text{eff}}} \mathcal{G}(\omega)$ still holds for \eqref{massless response fun}, guaranteed by the thermalization theorem \cite{sec1-7}.

For comparison, the response function for a detector coupled to a classical thermal bath at temperature $T_U=a / 2 \pi$ is: 
\begin{equation}
    \mathcal{G} _ {\text{thermal}}(\omega) \sim  \frac { 2 ^ { 2 - n } \pi ^ { \frac { 1 - n } { 2 } } \omega ^ { n - 3 } } { \Gamma ( \frac { n - 1 } { 2 } ) } \frac { 1 } {1- e ^ { - \beta \omega } }
    \label{thermal1}
\end{equation}
where no dependence on the spacetime dimension parity is present.


Inserting \eqref{massless response fun} and \eqref{thermal1} into \eqref{gamma}, one can obtain the Kossakowski coefficients for the different background fields as
\be
\left\{
\begin{aligned}
   \gamma_{+,~n}^{m=0}&= \mu^2\frac{ \pi^{\frac{n-5}{ 2}}\beta^{3-n}}{2\Gamma\left(\frac{n-1}{ 2}\right)}\left|\Gamma\left(\frac{n}{2}-1+ \frac{\beta\omega}{2\pi}i\right) \right|^{2} \operatorname{cosh}(\beta\omega / 2)    \\
     \gamma_{+,~n}^{\text{thermal}}&=\mu^2 \frac { 2 ^ { 2 - n } \pi ^ { \frac { 1 - n } { 2 } } } { \Gamma ( \frac { n - 1 } { 2 } ) } \frac {  \omega ^ { n - 3 } } {1- e ^ { - \beta \omega } }\left[1+(-1)^{n-2}e^{-\beta\omega}\right] 
\end{aligned}
\right.
\label{coefficients}
\ee
which, through the solution \eqref{solution}, determines the decay rate of the components of the detector's Bloch vector. 

A crucial insight from \eqref{coefficients} is that for the massless scalar field, the dependence of the detector response on spacetime dimension parity, evident in \eqref{massless response fun}, is absent in the decay rates governed by the Kossakowski coefficients. This explains why the statistics inversion phenomenon, while altering the detector excitation spectrum, does not affect the thermal interpretation of the Unruh effect \cite{sec1-7-9}\footnote{An alternative understanding of this point can be obtained from the perspective of open dynamics \cite{sec2-plus1}. Note that the response function is, in general, related to the early-time behavior of the detector's density matrix. Regardless of its complicated form (e.g., \eqref{massless response fun} depending on spacetime dimension parity), however, the master equation can resum this early-time behavior into a \emph{unique} late-time thermal state, which is guaranteed once the Wightman function satisfies the KMS condition.
}. However, a previously overlooked dependence on spacetime dimension parity is encoded in $\gamma_{+,~n}^{\text{thermal}}$. As we will demonstrate in Section \ref{sec4}, this parity dependence leads to an unexpected temporal behavior of the fidelity difference for the detector thermalization driven by a classical thermal field.

\section{Unruh thermalization process of the UDW detector}
\label{sec3}

The irreversible thermalization of the detector driven by the Unruh effect has a quantum origin. It is then legitimate for us to reexamine this process from a quantum thermodynamic perspective. End at a unique final state, the local dynamics of the detector, governed by the detector-field interaction, permit an ensemble of trajectories starting from different initial states \cite{Lidar}. In the following, we utilize trajectory-dependent thermodynamic process functions, especially quantum coherence and heat, which formulate the quantum First Law, to characterize the trajectories of the thermalization process. We define a measure on the trade-off between the detector's quantum coherence and heat, where its dependence on spacetime dimension distinguishes Unruh thermalization from a classical bath-driven thermalization.

\subsection{One-way trajectory in Bloch sphere}

\begin{figure}[htbp]
\centering
    \subfloat[Massless scalar]{\includegraphics[width=.42\textwidth]{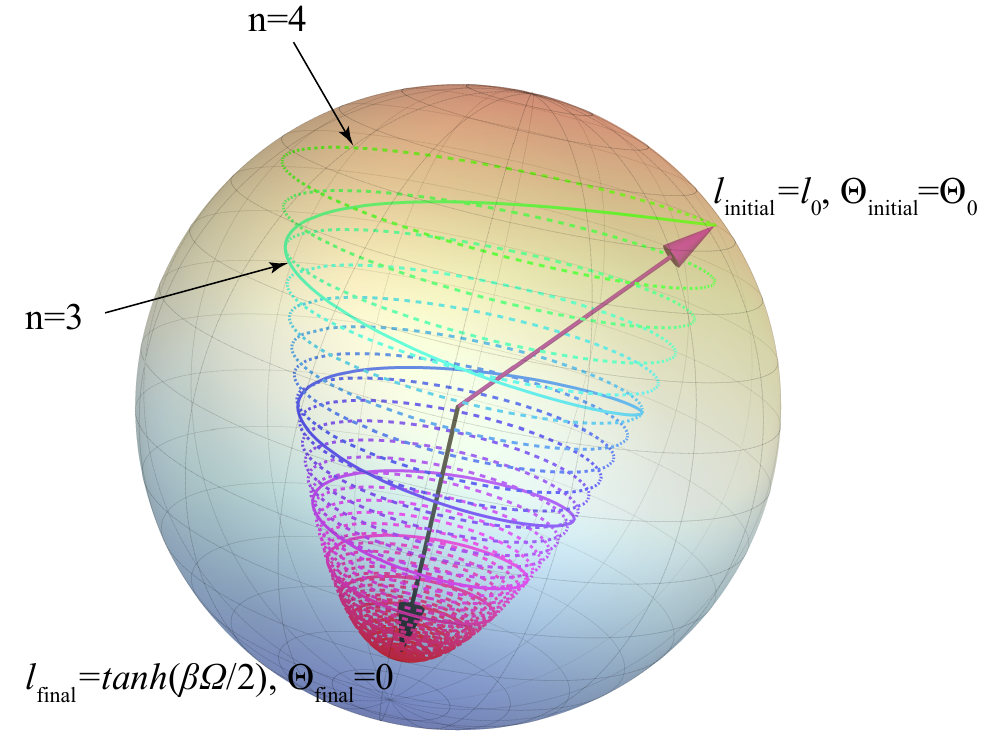}}~~~
    \subfloat[Thermal field]{\includegraphics[width=.43\textwidth]{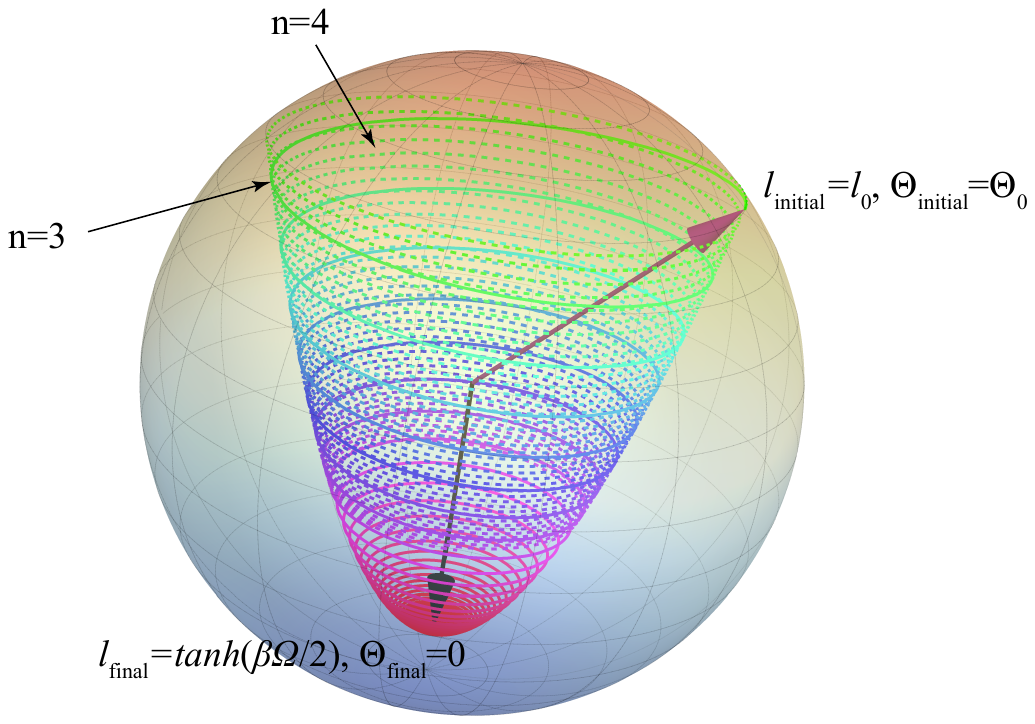}}
    \caption{The thermalization trajectories of the UDW detector in the Bloch sphere, when it interacts with (a) a massless scalar field, or (b) a classical thermal bath. The detector starts from a pure state characterized by $l_0=1,\Theta_0=\pi/4$. The estimation has been taken under $\beta=5$ and $\Omega=0.5$. The solid curves correspond to open dynamics in $n=3$-dim Minkowski spacetime, while dashed curves denote the thermalization trajectories of the detector moving in $n=4$-dim spacetime. Different trajectories ultimately converge at the same Gibbs state with the Bloch vector length $l_{\text{final}}=\tanh\left(\beta\Omega/2\right)$.}
    \label{Trajectory}
\end{figure}

It is instructive to visualize the thermalization trajectory of the detector in the Bloch sphere. Fig.\ref{Trajectory} illustrates the time evolution of the detector's Bloch vector, given by Eq.\eqref{solution}, for its interaction with massless scalar and classical thermal bath, respectively.

We observe that as time passes, the detector's Bloch vector $\bm{n}(t)$ (red arrow) traces a spiral trajectory with a continuously decreasing radius, which eventually converges to the Bloch vector $(0,0,\gamma)$ denoting the final Gibbs state (black arrow along the negative z-axis). 
These one-way, non-intersecting trajectories within the Bloch sphere provide a geometric illustration of the irreversibility of the detector's thermalization process, an inherent feature of the GKSL evolution governed by the Markovian master equation. In particular, the dynamical maps defined by \eqref{master} form a CPTP semigroup \cite{sec1-8} and are essentially non-invertible and contractive. This means that the GKSL evolution induces a loss of distinguishability between the system state and the late-time thermal state, as illustrated in Fig.\ref{Trajectory}, where different spiral (thermalization) trajectories converge to the same (thermal) fixed point $(0,0,\gamma)$.

The influence of spacetime dimensionality becomes apparent when comparing trajectories for different $n$. For a massless scalar background (Fig.\ref{Trajectory}(a)), the trajectory length in $n=4$ dimensional Minkowski spacetime is significantly longer than in $n=3$ dimensional spacetime. This suggests that in higher-dimensional spacetime, the Unruh thermalization process endures over a larger timescale. On the other hand, it is notable that, being driven by a classical thermal bath at the Unruh temperature, an inertial detector always evolves along a considerably long trajectory through state space (Fig.\ref{Trajectory}(b)), regardless of the spacetime dimensionality. Nevertheless, we cannot conclude that this can serve as a signature to distinguish thermal radiation from Unruh radiation, since there is a lack of quantification verification from the visualization.

 \subsection{Thermodynamic process functions}

From a thermodynamic perspective, the thermalization of the UDW detector involves changing its internal energy. For a closed system that cannot exchange any matter with its surrounding medium, work and heat are the only two forms of energy that can be transferred. This is the core of the classical First Law, stated as the conservation of energy. Extended to the quantum regime, the detector internal energy is defined as $U \equiv\langle H\rangle=\operatorname{tr}\{\rho H\}$ \cite{sec3-0-2}. The energetic contribution of quantum coherence undergoing open dynamics must be taken into account, which is an element absent in both classical and stochastic thermodynamics.

For an open quantum system, in the instantaneous basis $| x( t )\rangle$, the density matrix $\rho( t )=\sum_x p_x( t )|x( t )\rangle\langle x( t )|$ is not commutative with its Hamiltonian $H=\sum_{n=1}^D E_n|n\rangle\langle n|$. This is because $c_{n x}:=\langle n \mid x\rangle$ is, in general, not a constant during the evolution, or, equivalently, quantum coherence (off-diagonal terms) of $\rho(t)$ in the energy eigenbasis $\{|n(t)\rangle\}$ is generated during the evolution. The internal energy is then $U(t)=\text{Tr}\left[\rho(t)H\right]=\sum_{n, x} E_n p_x \left|c_{n x}\right|^2$, whose time derivative can be divided into three parts:
\be
\dot{U}(t)=\sum_{x, n}\left(\dot{E}_n p_x\left|c_{n x}\right|^2+E_n \dot{p}_x\left|c_{n x}\right|^2+E_n p_x\left|\dot{c}_{n x}\right|^2\right):=\dot{\mathbb{W}}+\dot{\mathbb{Q}}+\dot{\mathbb{C}}
\label{1stlaw}
\ee
Here, the first part $\dot{\mathbb W}\equiv \sum_{x, n}\dot{E}_n p_x\left|c_{n x}\right|^2$ is identified as the rate of quantum work, since it arises from changes in the energy levels $E_n(t)$ due to the driving of the Hamiltonian. The second part, $\dot{\mathbb{Q}}\equiv \sum_{x, n} E_n \dot{p}_x\left|c_{n x}\right|^2$, quantifies the internal energy change due to variations in the state population $\{p_x(t)\}$ (typically accompanied by entropy change due to environmental exchange), which is an extension of the classical heat change. The final part, $\dot{\mathbb C}\equiv \sum_{x, n} E_n p_x\left|\dot{c}_{n x}\right|^2$, arises from changes in the relative orientation between the eigenbases of $\rho(t)$ and $\hat H(t)$. This is a purely quantum-mechanical contribution due to the dynamics of coherence, as one can check that $\dot{\mathbb C}=0$ whenever $[\rho(t),\hat H(t)]=0$, which means the system has no energy-basis coherence. The formula Eq.\eqref{1stlaw} is the so-called First Law of quantum thermodynamics \cite{sec3-0-1}.

Specify to the UDW detector, we have $U_t:=\operatorname{Tr}\left[\rho(t) H_{\text {detector }}\right]=\frac{\omega}{2} l(t) \cos \Theta$, where $\cos \Theta=n_3 / l(t)$ is fixed by the angle between the Bloch vector and the $z$-axis. The time variation of three thermodynamic process functions (i.e., quantum work $\mathbb{W}$, heat $\mathbb{Q}$, and coherence $\mathbb{C}$) can be calculated straightforwardly as (for details see Appendix B of \cite{sec1-14})
\be
\left\{\begin{aligned}
\dot{\mathbb{W}}(t) & =\frac{l(t) \cos \Theta}{2} \frac{d \Omega}{d t}=0 \\
\dot{\mathbb{Q}}(t) & =\frac{\Omega \cos \Theta}{2} \frac{d l(t)}{d t}=-g^2 \Omega\left[\left(\frac{\gamma_+ n_3}{2}+\gamma_{-}\right) \cos^2 \Theta+\frac{\gamma_{+} n_3}{2} \right] \\
\dot{\mathbb{C}}(t) & =\frac{\Omega l(t)}{2} \frac{d \cos \Theta}{d t}=-g^2 \Omega\left(\frac{\gamma_+ n_3}{2}+\gamma_{-}\right) \sin ^2 \Theta
\end{aligned}\right.
\ee
Here, the change rate of quantum work is zero because the detector Hamiltonian is time-independent. This indicates that, upon undergoing irreversible thermalization, the quantum work performed by the detector remains constant.

\begin{figure}[htbp]
\centering
\subfloat[Massless scalar]{\includegraphics[width=.47\textwidth]{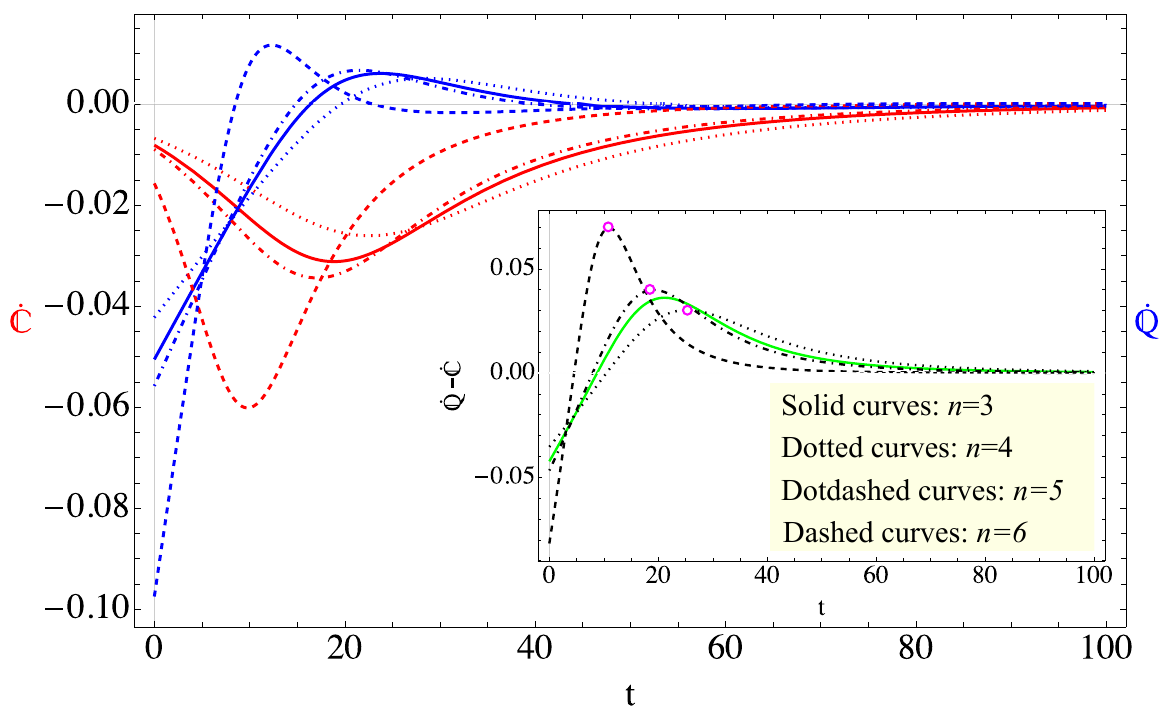}}~
    \subfloat[Thermal field]{\includegraphics[width=.49\textwidth]{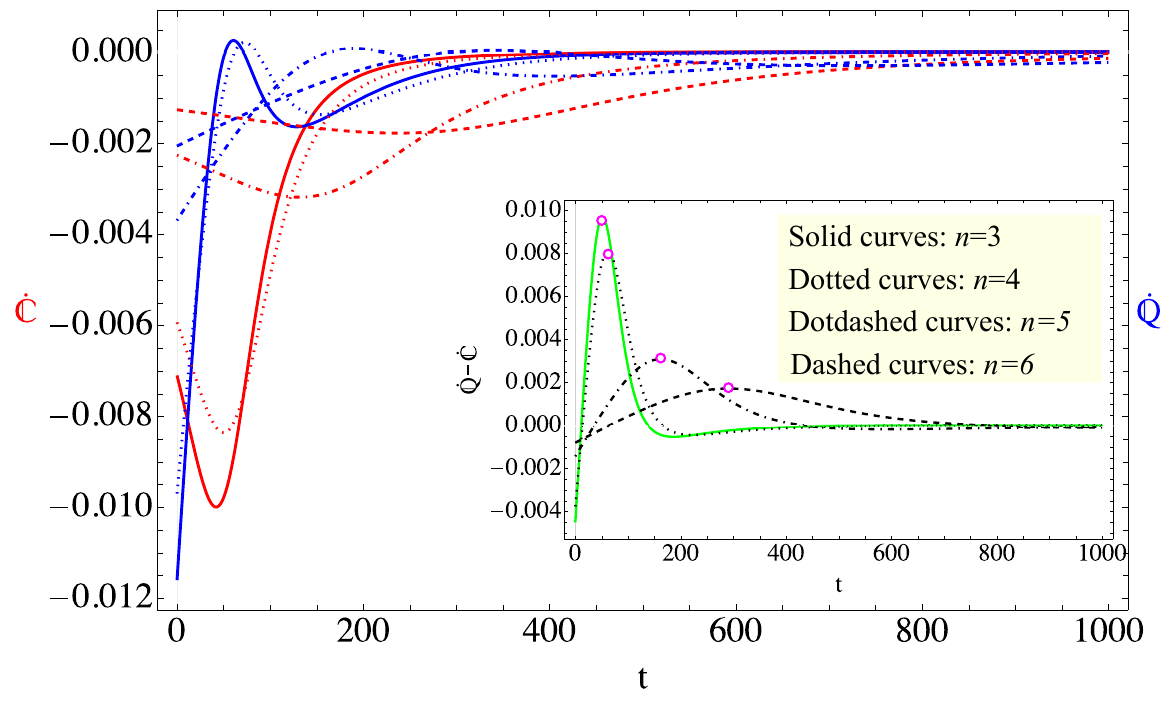}}
    \caption{The thermalization process characterized by the changing rate of quantum coherence and heat of the UDW detector, when it interacts with (a) a massless scalar field and (b) a classical thermal bath. The detector starts from a pure state characterized by $l_0=1,\Theta_0=\pi/4$. In the insets, the maximal value of a difference function $\Delta(t;n)$ between the change rates of quantum coherence and heat is located by pink circles for various spacetime dimensionality $n$. The estimation has been taken under $\beta=1$, $\Omega=2$. }
    \label{thermalfunc}
\end{figure}

We depict the change rates $\dot{\mathbb{Q}}$ and $\dot{\mathbb{C}}$ of two thermodynamic process functions, namely quantum coherence and heat of the detector. Given a fixed initial state ($l_0=1,	\Theta_0=\pi/4$), they depend on the detector's proper time and characterize a continuous thermalization trajectory that connects the initial state to a unique thermalized endpoint—a Gibbs state at temperature $\beta=2$. For comparison, we illustrate the scenario of Unruh thermalization of an accelerating detector with a massless scalar background (Fig.\ref{thermalfunc}(a)), as well as the scenario of thermalization driven by the classical bath for an inertial detector (Fig.\ref{thermalfunc}(b)).

In both cases, we see a complementary time evolution between the trajectory-dependent $\dot{\mathbb{Q}}$ (blue curves) and $\dot{\mathbb{C}}$ (red curves). This is expected from the quantum First Law \eqref{1stlaw} which, for the UDW detector model, claims that a combination of trajectory-dependent variations of quantum coherence and heat equals a state-dependent function $dU=\dbar{\mathbb{Q}}+\dbar{\mathbb{C}}$. Undergone the thermalization process, such complementary behavior between $\dot{\mathbb{Q}}$ and $\dot{\mathbb{C}}$ indicates a trade-off between quantum coherence and heat\footnote{Similar behavior between quantum coherecne and heat (in terms of the classical Kullback-Leibler divergence) was previously observed \cite{sec1-13} from Hawking thermalization process in Schwarzschild spacetime.}.

We observe two signatures that distinguish Unruh thermalization from the process driven by a thermal bath. First, after a sufficiently long time, the curves of change rates $\dot{\mathbb{Q}}$ and $\dot{\mathbb{C}}$ are convergent, which means the detector has approached an equilibrium. We note that the timescale of thermalization with a classical bath (e.g., $\sim\mathcal{O}(10^3)$ in Fig.\ref{thermalfunc}(b)) is significantly longer than the scale of the Unruh thermalization (e.g., $\sim\mathcal{O}(10^2)$ in Fig.\ref{thermalfunc}(a)). This is a qualification verification of what was previously observed in Fig.\ref{Trajectory}. 

Second, we define a difference between the change rates of quantum coherence and heat as $\Delta(t;n):=\dot{\mathbb{Q}}-\dot{\mathbb{C}}$ and illustrate it in the insets of Fig.\ref{thermalfunc}. We find its maximal value $\text{Max}_{t\in\mathbb{R}^+}\left(\Delta(t;n)\right)$ as a function of spacetime dimensionality behaves very differently for Unruh thermalization and the thermalization driven by a classical bath. In particular, for the thermalization driven by thermal radiation, with increasing spacetime dimensionality $n$, the value of $\text{Max}_{t\in\mathbb{R}^+}\left(\Delta(t;n)\right)$ degrades monotonously. On the contrary, for the Unruh thermalization, $\text{Max}_{t\in\mathbb{R}^+}\left(\Delta(t;n)\right)$ has an increasing maximum if spacetime dimension $n$ grows.

Unfortunately, due to an exceptional $n=3$ case (green curve in the inset of Fig.\ref{thermalfunc}(a)), the function $\text{Max}_{t\in\mathbb{R}^+}\left(\Delta(t;n)\right)$ may serve as a limited indicator that distinguishes the Unruh effect from its classical counterpart. The exception may be ascribed to the mathematical particularity of the Kossakowski coefficient \cite{sec1-7-8}, which is $\beta$-independent then. Nevertheless, for $n\geqslant 4$, the dependence of $\text{Max}_{t\in\mathbb{R}^+}\left(\Delta(t;n)\right)$ on the spacetime dimensionality could still be a compelling diagnosis of Unruh thermalization from its classical counterpart, thus unraveling its quantum origin.

\section{Asymmetry of Unruh thermalization}
\label{sec4}

The irreversible thermalization of the UDW detector manifests as a one-way trajectory on the Bloch sphere (Fig.\ref{Trajectory}), traced by the motion of its (instantaneous) state point in Hilbert space. To complement the geometric information encoded in process functions (Fig.\ref{thermalfunc}), it is legitimate to investigate further the kinematics of state point "flowing" along the trajectory. In the classical regime, anomalous paths between two desired states exist during out-of-equilibrium thermodynamic processes, exhibiting exotic behavior. A remarkable example is the Mpemba effect \cite{mpemba-1}, in which an initially hot system is quenched into a cold bath and reaches equilibrium faster than an initially cooler system. This counterintuitive phenomenon has recently been extended to the quantum regime and for open quantum systems \cite{add}.

In this section, we consider the UDW detector starting from an initial Gibbs state at an effective temperature $T_0$, and it evolves to the final thermal equilibrium at the Unruh temperature $T_U$. We demonstrate a quantum Mpemba-like phenomenon \cite{sec1-16} for the Unruh effect. The thermalization process due to the Unruh effect is asymmetric, meaning the detector inherently follows different trajectories depending on whether it undergoes heating (if $T_0 < T_U$) or cooling (if $T_0 > T_U$). Then, a general asymmetry of the thermalization processes under the Unruh effect, starting from initial detector states with the same "distance" to the final Gibbs state, is discussed.

\subsection{Information geometry and quantum thermal kinematics}
\label{sec4.1}

To proceed, we need to introduce the necessary tools from information geometry of quantum state space \cite{sec3-1}, especially the measures to quantify the "distance" and evolution "flow" between system states, which are considered as points within a geometric space. 

A commonly used measure of how close two quantum states are in terms of their density matrix is the Uhlmann fidelity
\be
F\left(\rho_1, \rho_2\right):=\left( \operatorname{Tr}\left[\sqrt{\sqrt{\rho_1}\rho_2\sqrt{\rho_1}}\right] \right)^2,
\label{eq3-1}
\ee
which is bounded $0\leqslant F\leqslant 1$. The fidelity is symmetric and invariant under unitary operations, but becomes asymmetric for an open quantum system due to the dissipator in the quantum master equation. Despite not defining a metric distance, the fidelity offers an easily computed statistical distance measure\footnote{The fidelity allows us to define the celebrated Bures distance $\left[D_B(\rho, \sigma)\right]^2:=2[1-F(\rho, \sigma)]$, whose infinitesimal line elment can be related to the QFI with respect to the parameter time \cite{sec3-plus-1} as $\left[D_B(\rho( t ), \rho( t +d  t ))\right]^2=\frac{1}{4} \mathcal{I}_Q[\rho( t )] d  t ^2+\mathcal{O}\left(d t^4\right)$, acquired the role of a symmetric and proper metric distance.}. Later, we regard the quantum states that share the same fidelity value to the specific thermalization end as being located at equal distances from it in the quantum state space. For example, the fidelity between the UDW detector \eqref{bloch} and its thermalization end \eqref{rho_equil} characterized by Unruh temperature $T_U$ can be straightforwardly calculated \cite{sec1-14} as
\be
F\left(\rho( t ), \rho_{\text{eq}}\right)=\frac{1}{2}\left\{1-n_3( t ) \tanh \left(\frac{\Omega}{2 T_U}\right)\right.
\left.+\sqrt{\left[1-l^2( t )\right] \operatorname{sech}^2\left(\frac{\Omega}{2 T_U}\right)}\right\},
\label{constr}
\ee 
which is not hard to be recast into
\be
{n_1^2+n_2^2}+\frac{\left[n_3-\left(1-2 F\right)\gamma_U\right]^2}{\left(1-\gamma_U^2\right)}=4 F(1-F).
\label{ellipsoid}
\ee
It manifests a $z$-axis symmetric ellipsoid centered at $(0,0,(1-2F)\gamma_U)$ in the quantum state space if taking a constant fidelity, where $\gamma_U:= \tanh \left({\Omega}/{2 T_U}\right)$. As time passes $ t \rar\infty$, the ellipsoid would eventually shrink to the point of $\rho_{\text{eq}}$. The Eq.\eqref{ellipsoid} indicates that at each moment, the detector states sharing the same "distance" to the given Gibbs state are distributed on a surface of an ellipsoid rather than on a smaller sphere inside the Bloch sphere. We infer that the submanifold of quantum states restricted by the detector open dynamics has a nontrivial geometry (metric) \cite{sec3-plus-2}. 
%

We need tools to quantify the kinematic details of the detector state along the trajectory. Working on the basis $| x( t )\rangle$ of the spectral decomposition $\rho( t )=\sum_x p_x( t )|x( t )\rangle\langle x( t )|$, the open dynamics described by the quantum master equation \eqref{master} can be regarded as a series of jump operators between instantaneous eigenstates of the detector. We define a quantum Fisher information (QFI) for the time parameter as
\be
\mathcal{I}_Q( t )=2\sum_{x, y} \frac{\left|\partial_ t  {\rho}_{x y}( t )\right|^2}{p_x( t ) +p_y( t ) },
\label{eq3-2}
\ee
where $\partial_ t  {\rho}_{x y}:=\langle x( t )| \partial_ t  {\rho}|y( t )\rangle$. The QFI can be decomposed as
\be
\mathcal{I}_Q(t ):=\mathcal{I}_Q^{incoh}+\mathcal{I}_Q^{coh} =\sum_x p_x(t)\left(\frac{d}{dt}\log p_x(t)\right)^2+2\sum_{x\neq y} \frac{\left|\partial_t {\rho}_{x y}(t)\right|^2}{p_x(t ) +p_y( t )}
\label{eq3-2-1}
\ee
where the first part, $\mathcal{I}_Q^{incoh}$, is the so-called incoherent QFI determined by eigenvalues $p_x$ and $p_y$, and $\mathcal{I}_Q^{coh}$ is the coherent QFI that depends on the off-diagonal components of the density matrix. In quantum state space, $\mathcal{I}_Q$ is identified as a contractive Riemannian metric under quantum stochastic maps \cite{sec3-5}, and plays a significant role in studying quantum (thermodynamic) evolution \cite{sec3-2}, as well as sets the bounds on quantum speed limits of the evolution \cite{sec3-3,sec3-4}.  

The infinitesimal metric distance is a squared line element mapped from the QFI by $ds^2=\frac{1}{4}\mathcal{I}_Qd t ^2$, which, along the system evolution path $\Gamma$ in quantum state space, gives the path length
\be
\mathcal{L}:=\int_\Gamma ds=\frac{1}{2}\int_0^T d t '\sqrt{\mathcal{I}_Q( t ')},
\label{eqT20}
\ee
that faithfully measures the distinguishability between the initial state $\rho(0)$ and the final state $\rho(T)$.

To study the kinematics of detector thermalization in the quantum regime, in particular, to quantify the temporal variation of local flows in quantum state space, we define two important measures induced from $\mathcal{I}_Q$ \cite{sec3-6}. First, we note that from the length \eqref{eqT20}, we can identify
\be
v_{Q}:=ds/d t =\frac{1}{2}\sqrt{\mathcal{I}_Q( t )},
\label{eqT35}
\ee 
as the instantaneous quantum "speed" of evolution. For a non-equilibrium quantum process, comparing the velocity $v_{Q}$ of the process and its reverse, the appearing asymmetry then characterizes thermodynamic irreversibility.

Another way to explore quantum kinematics is through the so-called quantum degree of completion (QDC), defined as the ratio of the length of the system's evolution path \eqref{eqT20} over different end times:
\be
\mathcal{R}_{ t /T}:=\frac{\mathcal{L}\left(0,  t \right)}{\mathcal{L}\left(0, T\right)},
\label{QDC}
\ee
which is a monotonically increasing function bounded between 0 and 1. This measure is especially useful for dissipative processes, where reaching a steady state through thermalization, i.e., traveling over a finite length \eqref{eqT20}, may take an infinite amount of time.

 \subsection{Quantum Mpemba-like effect}
 \label{sec4.2}

 \subsubsection{Two-temperature protocol} 
  \label{4.2.1}
 
We now examine a simple protocol to investigate the asymmetry of Unruh thermalization, constituted by a cooling and heating process of an accelerating UDW detector. 

\begin{figure}[htbp]
\centering
\subfloat[Two-temperature protocol]{\includegraphics[width=.45\textwidth]{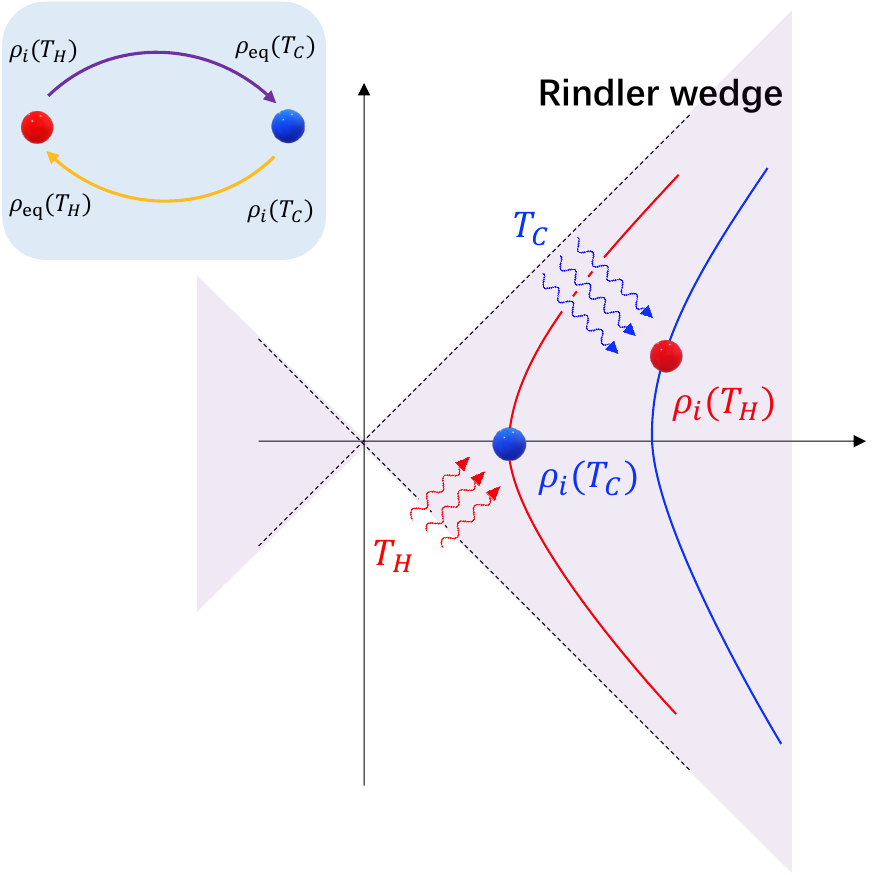}}~~~~~~~~
    \subfloat[Three-temperature protocol]{\includegraphics[width=.52\textwidth]{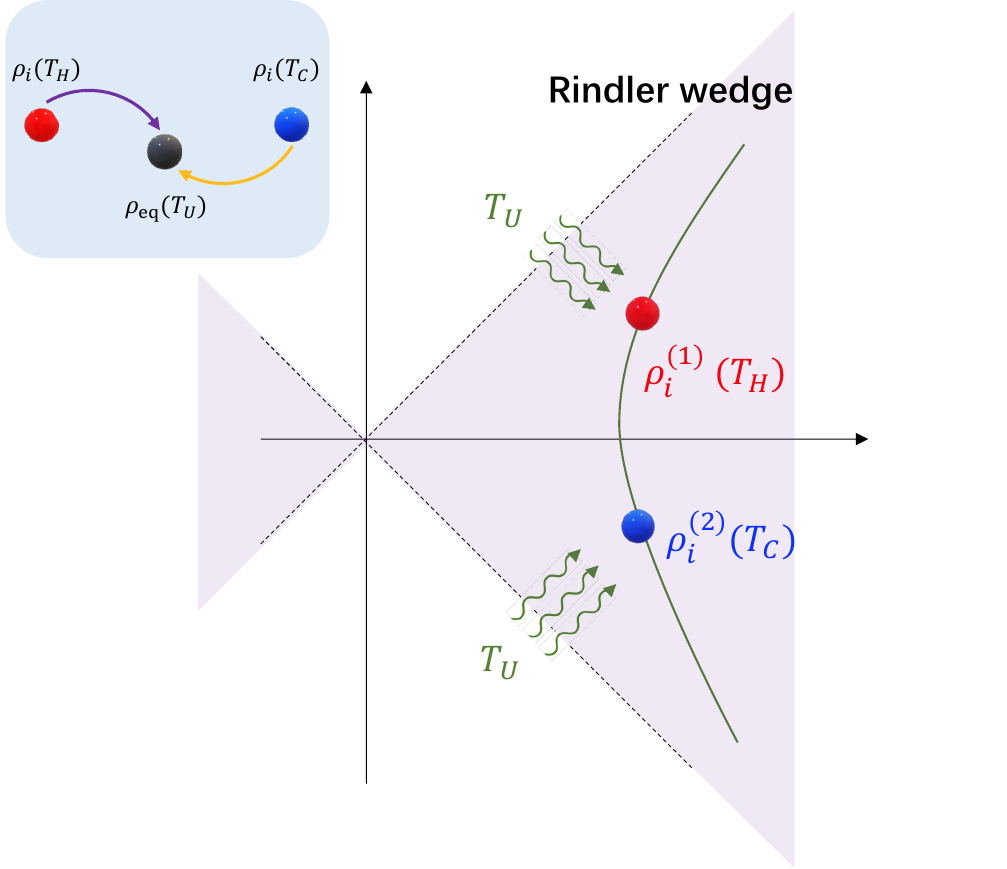}}
    \caption{The protocols examine the asymmetry of detector thermalization under the Unruh effect. In particular, (a) shows a two-temperature protocol demonstrating the detector's heating/cooling process, and (b) illustrates a three-temperature protocol where two detectors, simultaneously starting from a cold and hot state respectively, are thermalized to a Gibbs state at intermediate temperature $T_M$, which was initially equal-fidelity to the two detectors. }
    \label{Protocol}
\end{figure}

As demonstrated in Fig.\ref{Protocol}(a), in a heating process, we assume the detector is initially prepared in a Gibbs state
\be
\rho_{i, \text {cold}}=\frac{1}{2}\left(1-\tanh \left(\frac{\Omega}{2 T_C}\right) \sigma_3\right) 
\ee 
at a lower temperature $T_C$. After being quenched into Unruh radiation, the detector thermalizes to the final thermal state \eqref{rho_equil} at a higher temperature $T_H$. At any specific time during this heating process, the fidelity of the detector with respect to the final Gibbs state is
\be
F_{\text {heating}}(t)=\frac{1}{2}\left\{1-n_{3}\left( t;T_C \right)\tanh \left(\frac{\Omega}{2 T_H}\right)+\sqrt{\frac{1-n^2_3\left( t;T_C \right)}{\cosh^2 \left(\frac{\Omega}{2 T_H}\right)}}\right\}
\ee
where $n_{3}( t;T_C )$ is given by \eqref{solution}, with the initial conditions $l_0=\tanh \left(\Omega / 2 T_C\right)$ and $\Theta_0=\pi$ determined from the state $\rho_{i, \text {cold}}$.

A similar cooling process can be realized once the UDW detector is initiated from a state
\be
\rho_{i, \text {hot}}=\frac{1}{2}\left(1-\tanh \left(\frac{\omega}{2 T_H}\right) \sigma_3\right)
\ee 
with a higher temperature $T_H$. If the detector experiences a uniform acceleration causing a Unruh radiation at a lower temperature $T_C$, it will eventually be cooled to the final equilibrium state \eqref{rho_equil} at $T_C$. The fidelity corresponding to this cooling process is
\be
F_{\text {cooling}}(t)=\frac{1}{2}\left\{1-n_{3}\left( t;T_H \right)\tanh \left(\frac{\Omega}{2 T_C}\right)+\sqrt{\frac{1-n_3^2\left( t;T_H \right)}{\cosh^2 \left(\frac{\Omega}{2 T_C}\right)}}\right\}
\ee
where $n_{3}( t;T_H )$ is given with the initial conditions $l_0=\tanh \left(\Omega / 2 T_H\right)$ and $\Theta_0=\pi$ determined by the hot state $\rho_{i, \text {hot}}$.

\begin{figure}[htbp]
\centering
    \subfloat[Massless scalar]{\includegraphics[width=.48\textwidth]{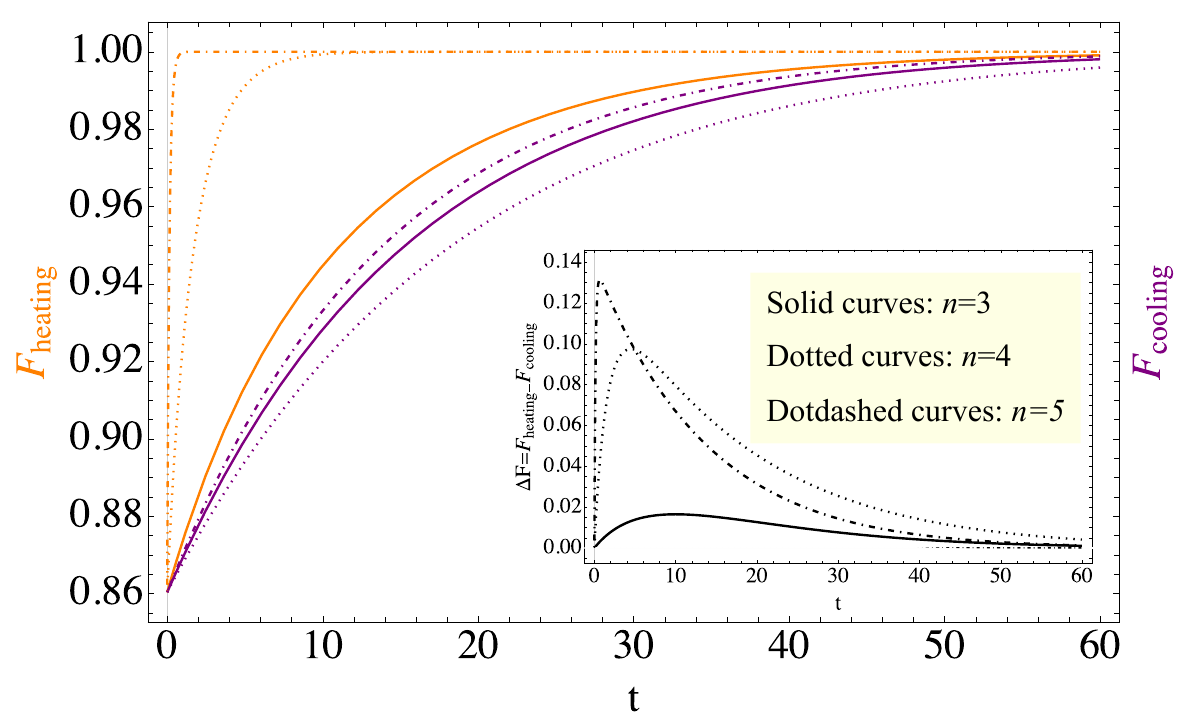}}~
    \subfloat[Thermal field]{\includegraphics[width=.48\textwidth]{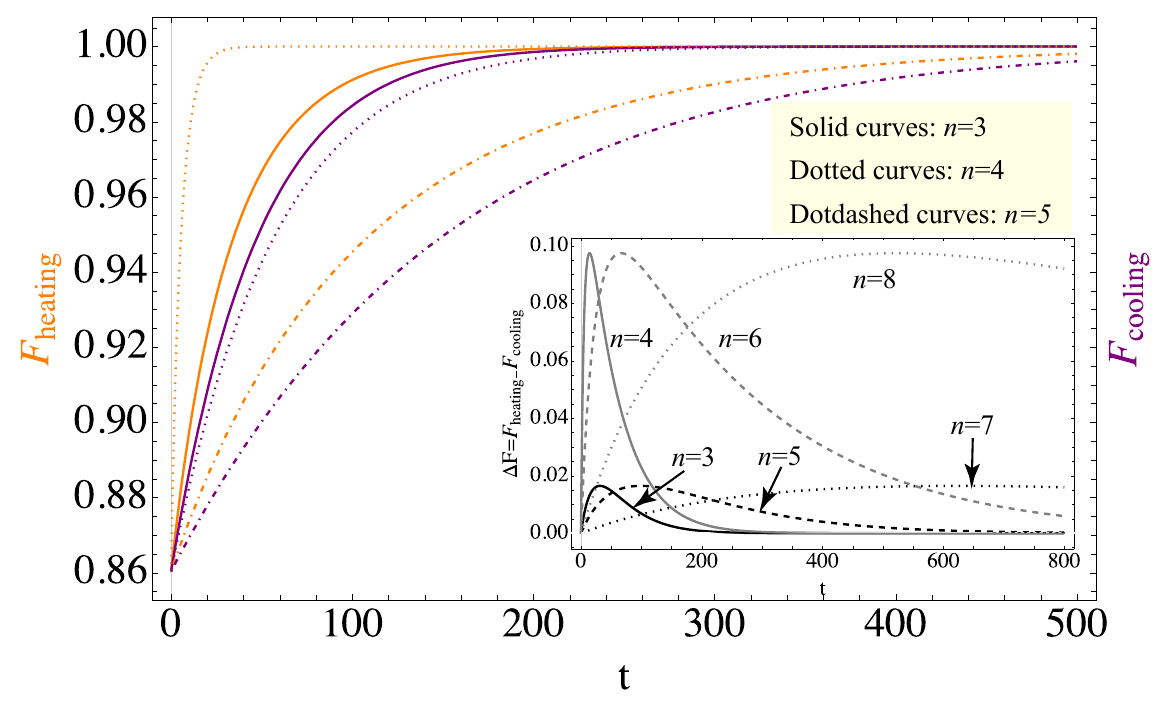}}
    \caption{The fidelity for the heating and cooling processes of a UDW detector in Minkowski spacetime. The detector heats up (cools down) to a higher (lower) temperature $T_H$ ($T_C$), starting from an initial Gibbs state at a lower (higher) temperature $T_C$ ($T_H$). The fidelity for the heating process (orange) consistently surpasses that of the cooling process (purple) across all scenarios involving interaction with (a) a massless scalar field, or (b) a classical thermal bath. The estimation was performed at $\Omega=2$, $T_C=1$, and $T_H=10$. In the inset, the maxima of the process fidelity difference $\Delta F=F_{\text {heating}}-F_{\text {cooling}}$ with respect to growing spacetime dimension are illustrated.}
    \label{Fidelity}
\end{figure}

For an accelerating UDW detector coupling to a massless scalar field, we plot the fidelity over time during the heating and cooling processes between two fixed temperatures, $T_C$ and $T_H$, in Fig.\ref{Fidelity}(a). For comparison, we also show the fidelity evolution for an inertial detector undergoing thermalization driven by a classical thermal bath at fixed temperatures $T_C$ or $T_H$ (see Fig.\ref{Fidelity}(b)). To refine the comparison, we also depict the fidelity difference $\Delta F:=F_{\text {heating}}-F_{\text {cooling}}$ in the insets.

For the massless scalar, the detector undergoes Unruh thermalization. We observe that the fidelity for the heating process $F_{\text {heating}}$ (orange curves) consistently surpasses that of the cooling process $F_{\text {cooling}}$ (purple curves) for arbitrary spacetime dimensions. Since fidelity measures the distance between quantum states, Fig.\ref{Fidelity}(a) indicates that the detector approach to its thermalization end is faster during its heating process than in its cooling process. We refer to the phenomenon as the quantum Mpemba-like effect (QME) \cite{Mpemba0,sec1-16,Mpemba1} under Unruh thermalization\footnote{We don't call it a genuine Mpemba effect because strictly speaking, the latter is additionally characterized by an exponential acceleration of the heating or cooling \cite{add}.}. In the inset, we show that the maxima of the fidelity difference $\Delta F$ increase monotonically with the growing spacetime dimension for Unruh thermalization. In Fig.\ref{Fidelity}(b), the fidelity of an inertial detector undergoing classical bath-driven thermalization is depicted. Besides a similar QME, we observe that the timescale for fidelity convergence is significantly longer than in the case of Unruh thermalization, which supports previous observations from thermodynamic process functions in Fig.\ref{thermalfunc}.


A surprise emerges when examining a Mpemba-like phenomenon \cite{sec3-6} for an inertial detector interacting with a classical thermal bath. As shown in the inset of Fig.\ref{Fidelity}(b), we observe that the maximum of the fidelity difference $\Delta F$ remains constant for the thermal bath in even-dimensional spacetime, but it has a lower constant value for the bath in odd-dimensional spacetime. While the heating process is still faster, this anomalous dependence of the fidelity difference on the parity of spacetime dimension suggests that one cannot equate the Unruh effect in an accelerated frame to an inertial thermal bath at the same temperature, even if the UDW detector exhibits the same Planckian response spectrum asymptotically. From a practical perspective, this suggests to us that the fidelity difference $\Delta F$ can be regarded as a signature to distinguish the Unruh effect from its classical counterpart.

To obtain additional insight into the quantum kinematics of Unruh thermalization, we use the $\mathcal{I}_Q$ to compute the instantaneous quantum "speed" \eqref{eqT35} and thermal kinematic distance, i.e., QDC \eqref{eq3-2}, for the two-temperature protocol.

By the definition of $\mathcal{I}_Q$, we need to work in a eigenbasis of the detector density matrix \eqref{solution}, which  can then be diagonalized as $\rho( t )=p_{+}|+\rangle\langle+|+p_{-}|-\rangle\langle-|$, where the eigenvalues $p_{ \pm}( t )=\frac{1}{2}(1 \pm l( t ))$ and eigenvectors are (up to a phase $\varphi_ t $):
\be
|+\rangle=\left(e^{-i \varphi_t} \cos \frac{\Theta}{2}, \sin \frac{\Theta}{2}\right)^{\top},~~~~~
|-\rangle=\left(-\sin \frac{\Theta}{2}, e^{i \varphi_t} \cos \frac{\Theta}{2}\right)^{\top}
\label{eigenbasis}
\ee
Here, $\tan{\varphi_t}=n_2/n_1$ and $\Theta$ is the angle of detector Bloch vector to $z$-axis, satisfying $\cos \Theta=n_3 / l( t )$. Inserting \eqref{solution} and \eqref{eigenbasis} into the definition \eqref{eq3-2}, we obtain the QFI with respect to the time parameter $ t $ as:
\be
\mathcal{I}_Q=\mathcal{I}_Q^{incoh}+\mathcal{I}_Q^{coh}=\frac{4\left\{\gamma_{-} \cos \Theta+  \gamma_{+}l\left(1-\frac{1}{2} \sin ^2 \Theta\right)\right\}^2}{1-l^2} +\Omega^2 l^2  \sin ^2 \Theta
\label{QFI2t}
\ee
where we have decomposed the QFI into a combination of the first $\mathcal{I}_Q^{incoh}$ terms and the second $\mathcal{I}_Q^{coh}$ term, respectively. The QFI then determines the instantaneous velocity $v_Q=\sqrt{\mathcal{I}_Q( t )}/2$ and the QDC at a specific time $\mathcal{R}_{ t _0/T}$.

\begin{figure}[htbp]
\centering
    \subfloat[Massless scalar]{\includegraphics[width=.48\textwidth]{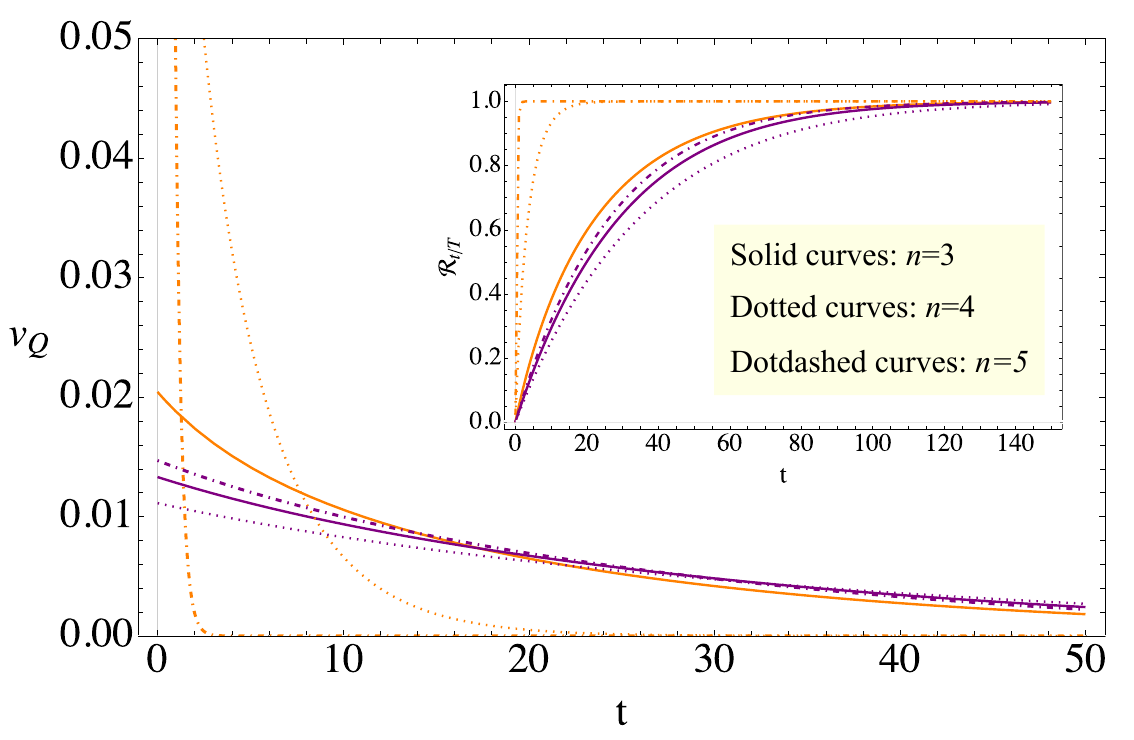}}
        \subfloat[Thermal field]{\includegraphics[width=.48\textwidth]{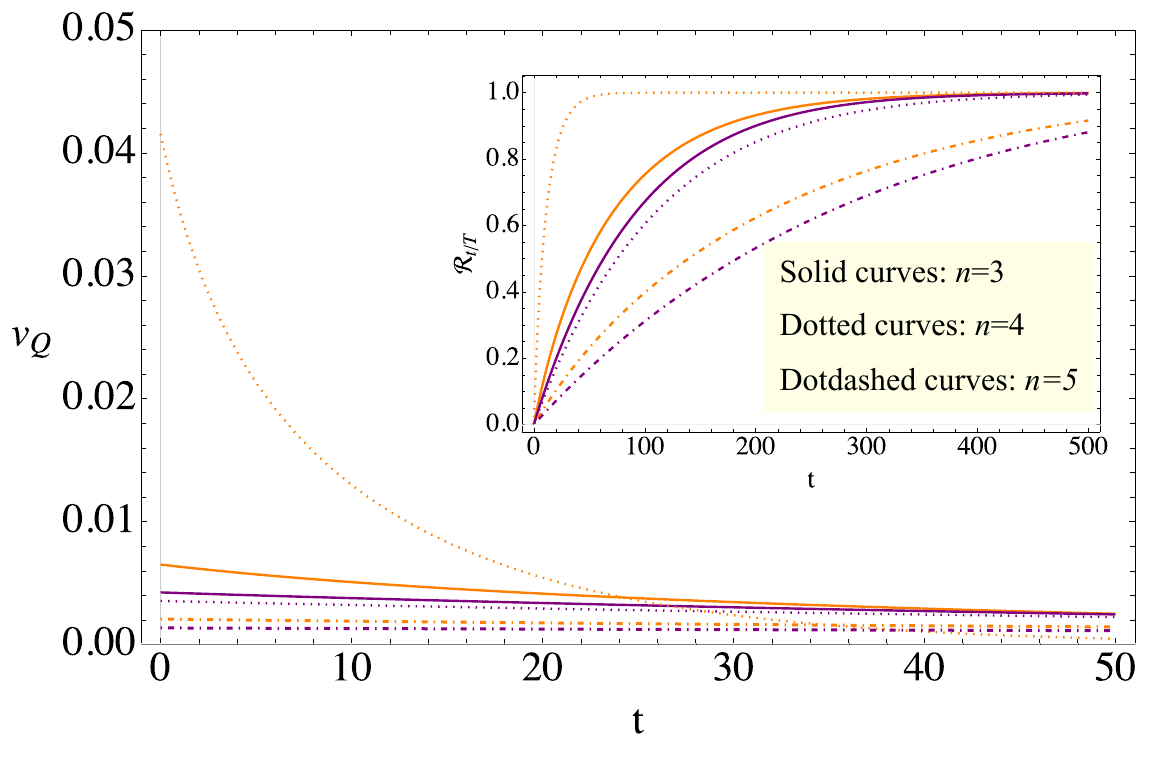}}
    \caption{The velocity for the heating and cooling processes of a UDW detector between a pair of temperatures $(T_C=1,T_H=10)$, once interacting with (a) a massless scalar, or (b) a classical thermal bath in Minkowski spacetime. While the cooling velocity crosses the heating one, the QDC of detector heating always exceeds the QDC of cooling. The estimation was performed at $\Omega=2$.}
    \label{velocity-QDC}
\end{figure}

%
%

We are now in a position to explore the quantum thermal kinematics by depicting the quantum instantaneous velocity of the UDW detector interacting with a massless scalar field (Fig.\ref{velocity-QDC}(a)) and a classical thermal bath (Fig.\ref{velocity-QDC}(b)), respectively. We observe that at the beginning of the thermalization process, the detector heats up at a faster rate, which consistently reduces and then crosses over the cooling velocity after a specific time. Nevertheless, the QDC of the heating process always exceeds that of the cooling protocol (insets of Fig.\ref{velocity-QDC}(a)(b)), indicating that toward the same end state, the heating detector always completes its thermalization faster than the cooling detector. On the other hand, the longer thermalization timescale for a classical bath driving can be observed by comparing the QDC convergent timescales in Fig.\ref{velocity-QDC}(b) and Fig.\ref{velocity-QDC}(a).

Finally, we note that since we examine a dissipative process between a pair of Gibbs states, the angle between the initial detector state and the $z$-axis is fixed as $\Theta_0=\pi$. This means that the quantum thermal kinematics (velocity and QDC) of the detector thermalization is determined solely by $\mathcal{I}_Q^{incoh}$, the incoherent part of QFI \eqref{QFI2t}. In Section \ref{sec4.3}, we will discuss the quantum thermal kinematics of Unruh thermalization, starting with non-thermal states, where the contribution of $\mathcal{I}_Q^{coh}$, the coherent part of QFI, should be taken into account.

 \subsubsection{Three-temperature protocol} 
  \label{4.2.2}
  
We now move to an alternative heating/cooling protocol involving two UDW detectors (Fig.\ref{Protocol}(b)) and examine whether similar QME can be exhibited. 

Considering two detectors are prepared in Gibbs states at temperatures $T_H$ and $T_C$, respectively. Through Unruh thermalization\footnote{Essentially, what we compare are the distinct thermalization processes of the UDW detector for chosen pairs of initial and final states. Therefore, we assume that even if two detectors are simultaneously coupled to the scalar field, their dynamics cannot affect each other. This differs from the so-called entanglement-harvesting protocol \cite{rev-4}, where entanglement can be generated at late times due to indirect correlations between two detectors via a common field coupling.}, they ultimately evolve into the same equilibrium at a temperature $T_U$. We require three temperatures to satisfy
\be
F\left[\rho^{(1)}_i(T_H;t=0), \rho_{\mathrm{eq}}(T_U)\right]=F\left[\rho^{(2)}_i(T_C;t=0), \rho_{\mathrm{eq}}(T_U)\right],
\label{cond}
\ee
to ensure that two detectors are initially at equal distance from the thermalized end state, as measured by Uhlmann fidelity. 

In general, all equal-distance initial states must lie on the surface of the ellipsoid \eqref{ellipsoid}, which, combined with a general expression $(n_1=l_0\sin\Theta_0,0,n_3=l_0\cos\Theta_0)$ given by the open dynamics \eqref{solution}, can be rewritten as
\be
l^2_0\left(1-\sin^2\Theta_0\gamma_U^2\right)-2l_0\cos\Theta_0(1-2F)\gamma_U=f(\gamma_U,F)
\label{3T}
\ee 
Here, $f(\gamma_U,F):=4F(1-F)(1-\gamma^2_U)-(1-2F)^2\gamma_U^2$ is a detector-independent function.

\begin{figure}[htbp]
\centering
    \subfloat[Thermal initial states]{\includegraphics[width=.6\textwidth]{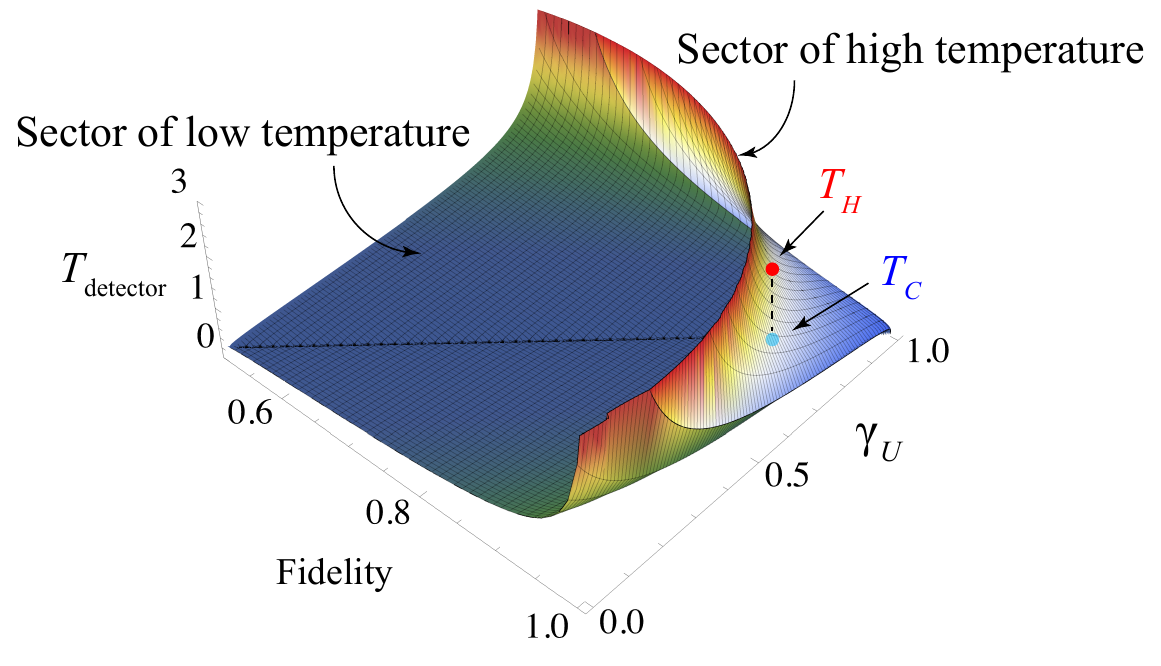}}~
    \subfloat[Nonthermal initial states]{\includegraphics[width=.35\textwidth]{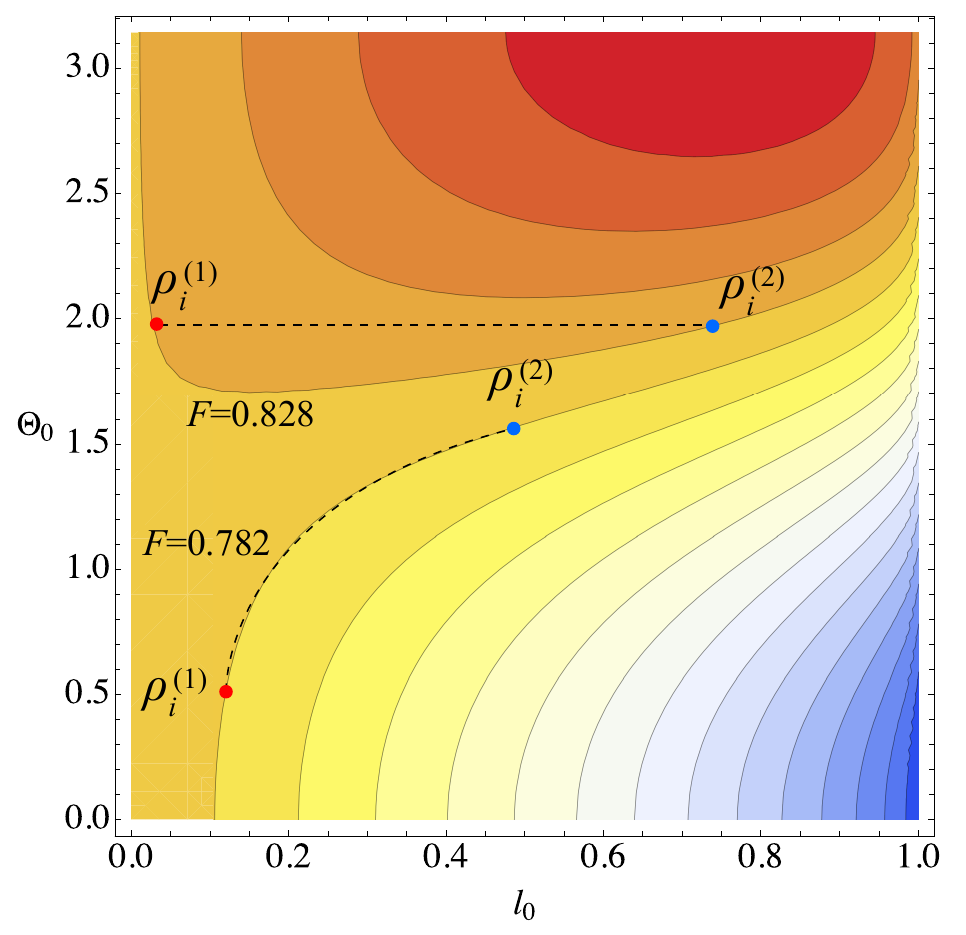}}
    \caption{Numerical illustration of the selection of equally distant initial states of two UDW detectors. (a) For the three-temperature protocol, two distinct sectors of allowed temperature regions are defined by \eqref{3T}, where a pair of temperatures $(T_H, T_C)$ can be chosen for specific values of fidelity and Unruh temperature. (b) For detectors starting from non-thermal states, two types of equally distant initial state pairs are present. The estimation was performed with $\Omega=1$, and the contours in (b) indicate constant fidelity when the massless scalar bath induces Unruh temperature $T_U=1/2$.  
    }
    \label{3T-solution}
\end{figure}

To conduct the three-temperature protocol, we prepare the two UDW detectors in equal-distant Gibbs states, which means $\Theta_0=\pi$ and $l_0=\tanh \left(\Omega / 2 T_{\text{detector}}\right)$. Substituting them into \eqref{3T}, we yield two distinct solutions as:
\be
T_{\text{detector}}=\frac{\Omega}{2\tanh^{-1}\left[(2F-1)\gamma_U\pm 2\sqrt{F(1-F)(1-\gamma^2_U)}\right]}.
\ee
Numerical analysis reveals that two distinct sectors of allowed temperature regions can only exist for restricted values\footnote{Besides being restricted in the domain of the denominator function, $F$ and $\gamma_U$ are also bounded by $F-\left(1-n_3\gamma_U\right)/2 \geqslant 0$. The constraint arising from \eqref{constr} indicates that equal-distance states cannot occupy the entire ellipsoid surface.} of $F$ and $\gamma_U$,  as shown in Fig.\ref{3T-solution}(a).

\begin{figure}[htbp]
\centering
    \subfloat[Massless scalar]{\includegraphics[width=.48\textwidth]{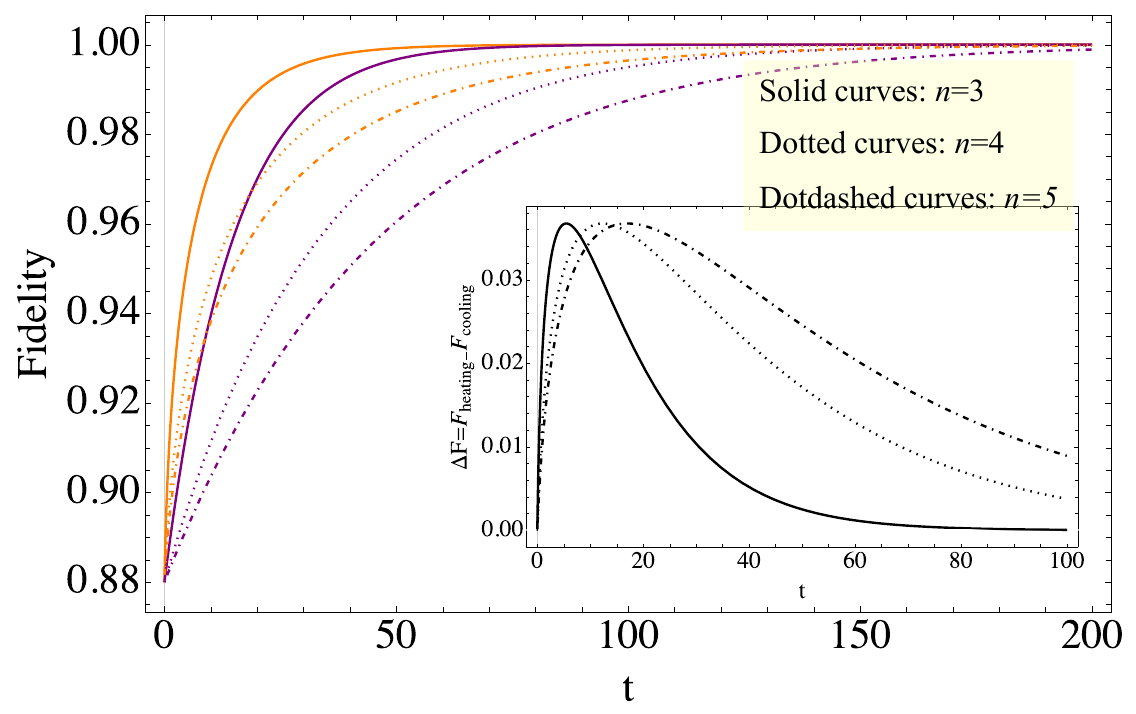}}
        \subfloat[Thermal field]{\includegraphics[width=.48\textwidth]{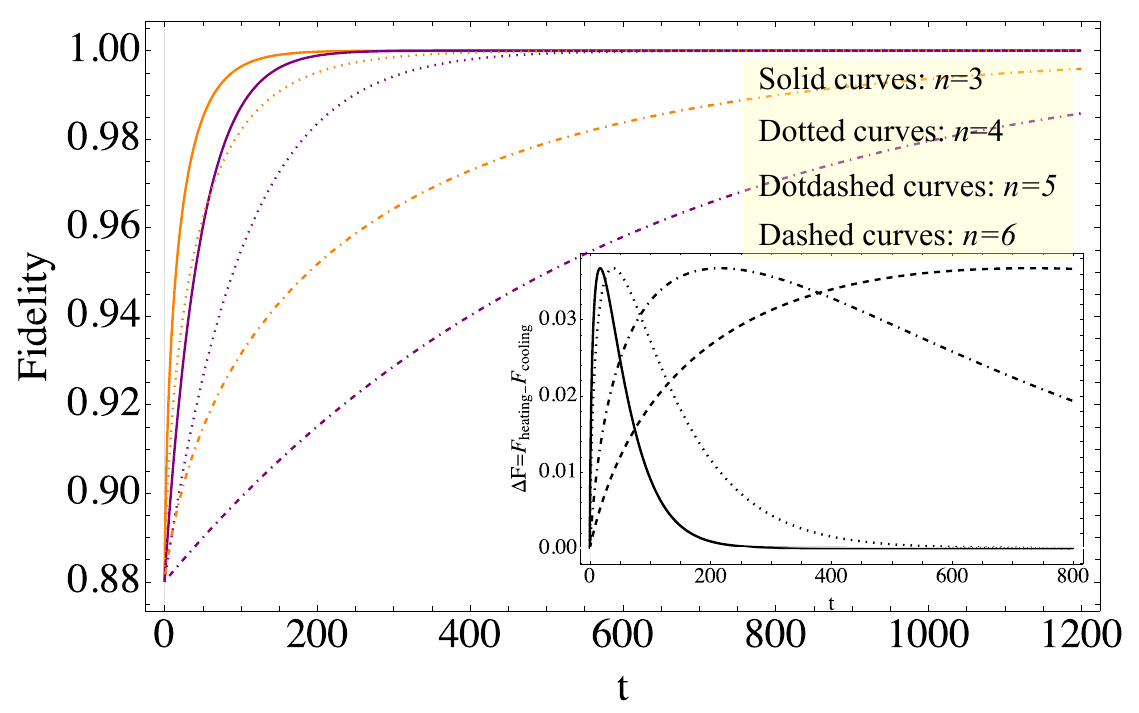}}
    \caption{The time-evolution of fidelity for the heating and cooling protocol of two UDW detectors interacting with (a) a massless scalar, or (b) a classical thermal bath in Minkowski spacetime. Initially, two detectors start from Gibbs states with temperatures $T_H=5.487$ and $T_C=0.146$, and they have equal "distance" $F=0.88$ to the same thermalization end at Unruh temperature $T_U=0.556$. As time passes, the fidelity of process $T_C\rar T_U$ (orange curves) always surpasses the fidelity of process $T_H\rar T_U$ (purple curves). 
    }
    \label{3T-fidelity}
\end{figure}

Without loss of generality, we demonstrate a numerical example where two UDW detectors are initially prepared in Gibbs states at temperatures $T_H=5.487$ and $T_C=0.146$, which have equal distances $F=0.88$ to the thermalization end at Unruh temperature $T_U=0.556$. The fidelity of two detectors $F\left[\rho^{(1)}_i(T_H;t), \rho_{\mathrm{eq}}(T_U)\right]$ and $F\left[\rho^{(2)}_i(T_C;t), \rho_{\mathrm{eq}}(T_U)\right]$, are time-dependent functions with the same boundary, starting from $F=0.88$ and converging to $F=1$. We illustrate the estimation results of the fidelity functions in Fig.\ref{3T-fidelity} for Unruh thermalization and classical bath-driven thermalization, respectively.

Compared to the two-temperature protocol, we observe distinct behaviors in the fidelity time evolution of the three-temperature protocol. The two detectors start from initial Gibbs states $\rho^{(1)}_i(T_H)$ and $\rho^{(2)}_i(T_C)$. As time passes, the fidelity during the heating process with $T_C\rar T_U$ (orange curves) consistently exceeds that during cooling with $T_H\rar T_U$ (purple curves). This indicates the QME in the three-temperature protocol, meaning the detector can heat up faster than it cools down to the same temperature (see Fig.\ref{3T-fidelity}(a)(b)). We observe that the spacetime dimensionality plays a different role compared to the two-temperature protocol. In particular, in higher-dimensional Minkowski spacetime, the fidelity curves for heating and cooling converge at a later time, which suggests a stretched timescale for the thermalization process. 

In the insets of Fig.\ref{3T-fidelity}(a)(b), we show the fidelity difference $\Delta F=F_{\text{heating}}-F_{\text{cooling}}$, which has the same maximum value regardless of the spacetime dimensionality. This quite differs from the behavior observed in the two-temperature protocol. We also note that in higher-dimensional spacetimes, the fidelity difference achieves its maximum and approaches zero at a later time. This supports the idea that detector heating/cooling under thermalization occurs over a stretched timescale as the spacetime dimensionality increases.

Unfortunately, we find that in the three-temperature protocol, the fidelity difference function $\Delta F$, which exhibits only minor numerical differences, fails to serve as a reliable discriminator between Unruh and classical bath-induced thermalization. Specifically, the dependency of the maximum $\Delta F$ on the parity of the spacetime dimension is absent. Nevertheless, compared to Unruh thermalization, a much longer thermalization timescale for classical bath driving is still observed, just as seen in the two-temperature protocol.

\begin{figure}[htbp]
\centering
    \subfloat[Massless scalar]{\includegraphics[width=.48\textwidth]{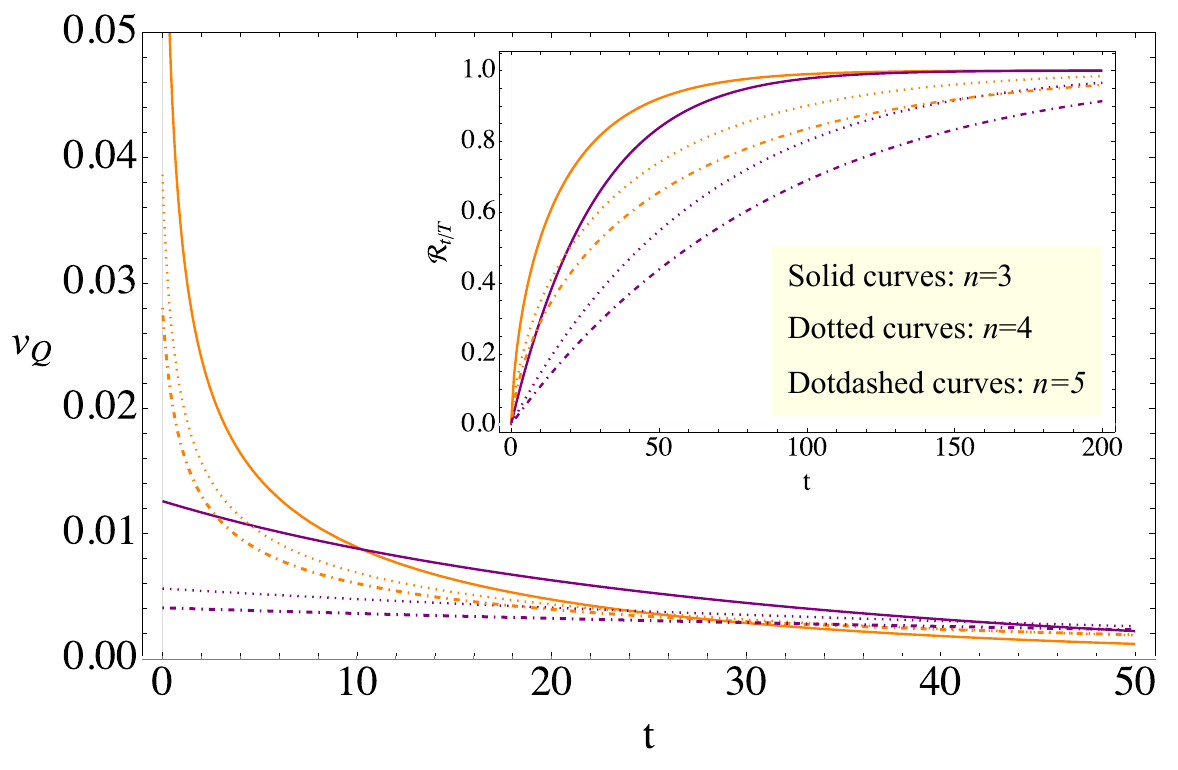}}
        \subfloat[Thermal field]{\includegraphics[width=.48\textwidth]{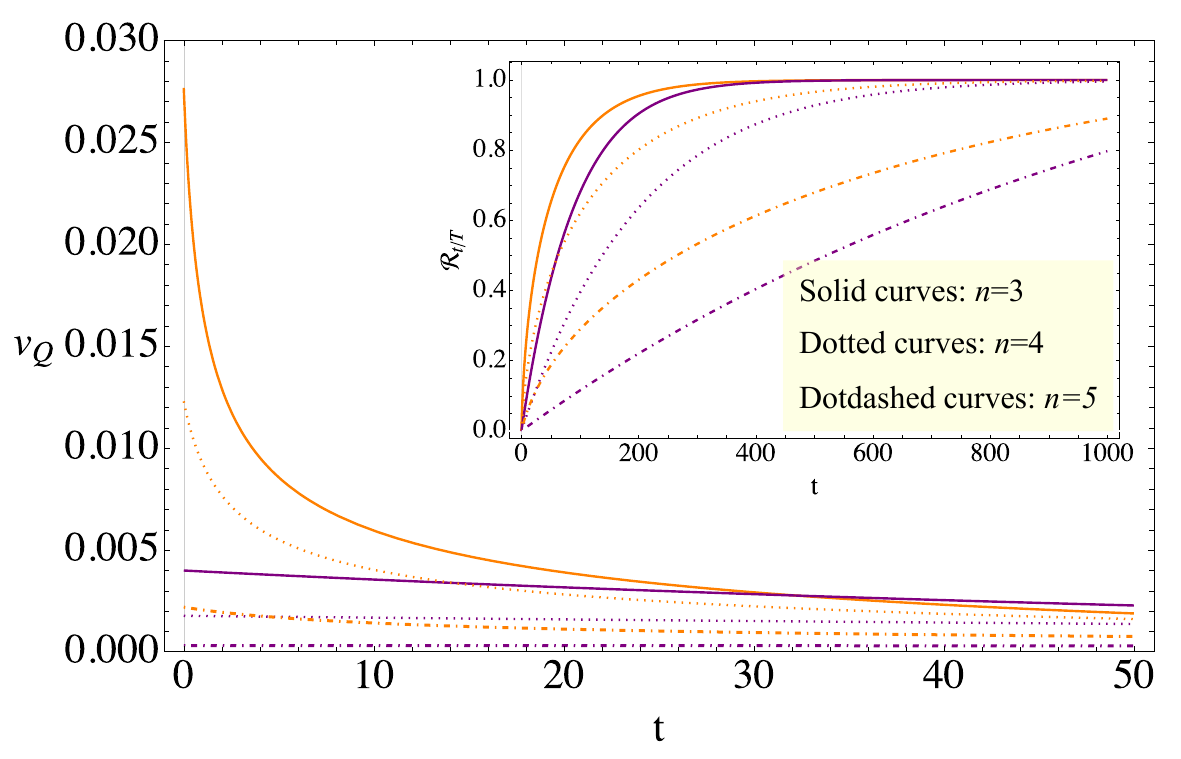}}
    \caption{The time-evolution of the velocity and QDC for the heating and cooling protocol of two UDW detectors interacting with (a) a massless scalar, or (b) a classical thermal bath in Minkowski spacetime. Initially, two detectors start from Gibbs states with temperatures $T_H=5.487$ and $T_C=0.146$, and they have equal "distance" $F=0.88$ to the same thermalization end at Unruh temperature $T_U=0.556$.
    }
    \label{3T-velocity-QDC}
\end{figure}


To complete the discussion, we depict the speed $v_Q$ as well as the QDC for the heating/cooling process under Unruh thermalization in Fig.\ref{3T-velocity-QDC}(a) and undergoing a classical bath-driven thermalization in Fig.\ref{3T-velocity-QDC}(b). As the spacetime dimension increases, the QDC of heating (orange curves) and cooling (purple curves) converge at a later time, indicating a longer timescale to reach the final thermalization end. As observed from the inset of Fig.\ref{3T-velocity-QDC}(b), the thermalization timescale for the classical bath is significantly longer than the timescale of Unruh thermalization driven by the scalar background.

\subsection{Unruh thermalization starting with non-thermal states}
\label{sec4.3}

While most equal-distant states on the ellipsoid surface \eqref{ellipsoid} are not thermal, it is legitimate to ask whether the asymmetry of thermalization observed before is universal for the detector starting from these non-thermal states. In particular, we compare the Unruh thermalization process of two UDW detectors, where one undergoes evolution $\rho_i^{(1)}(l_{0}^{(1)},\Theta_{0}^{(1)};t=0) \rar \rho_{\text{eq}}(T_U)$ and the other is thermalized as $\rho_i^{(2)}(l_{0}^{(2)},\Theta_{0}^{(2)};t=0) \rar \rho_{\text{eq}}(T_U)$. Similar to the three-temperature protocol (Fig.\ref{Protocol}(b)), we require that initially two detectors are equal-distant in the sense of
\be
F\left[\rho_i^{(1)}(l_{0}^{(1)},\Theta_{0}^{(1)};t=0), \rho_{\mathrm{eq}}(T_U)\right]=F\left[\rho_i^{(2)}(l_{0}^{(2)},\Theta_{0}^{(2)};t=0), \rho_{\mathrm{eq}}(T_U)\right].
\ee
We can specify the detector's initial states by solving \eqref{3T} with $\Theta_0\neq n\pi$ ($n\in\mathbb{Z}$). However, the undetermined parameter $\Theta_0$ makes even the numerical analysis much more involved than the three-temperature protocol. 

A more efficient approach is to identify the possible choices of initial state by observing the constant fidelity contours for varying $l_0$ and $\Theta_0$. For instance, we plot in Fig.\ref{3T-solution}(b) the fidelity between an arbitrary initial state of the detector and its unique thermalization end, undergoing Unruh thermalization driven in a massless scalar field background. The contour lines illustrate isofidelity curves. One can examine that for varying Unruh temperatures, pairs of equal-distant non-thermal states ($\rho_i^{(1)},\rho_i^{(2)}$) manifest exclusively in two distinct configurations\footnote{The possibility of a pair of equal-distant non-thermal states with the same $l_0$ but different $\Theta_0$ occurs when $T_U\rar\infty$. In fact, one can see that as $T_U\rar\infty$, Eq.\eqref{3T} reduces to an equation independent of $\Theta_0$. Since this is a physically uninteresting case, we will no longer consider it.}: (1) those with the same $\Theta_0$ but different $l_0$, geometrically, are possible when the whole ellipsoid is located at the negative $z$-axis; (2) those with both $\Theta_0$ and $l_0$ different. In the following, we examine the QME for these configurations by numerical examples.

\begin{figure}[htbp]
\centering
    \subfloat[Massless scalar]{\includegraphics[width=.49\textwidth]{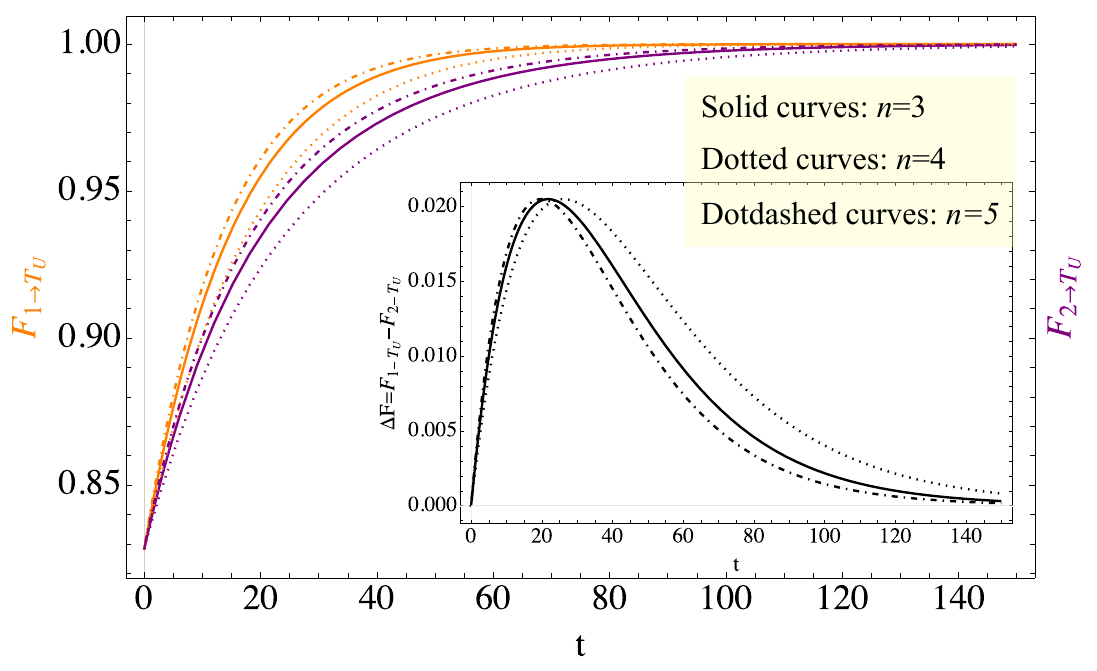}}
        \subfloat[Thermal field]{\includegraphics[width=.49\textwidth]{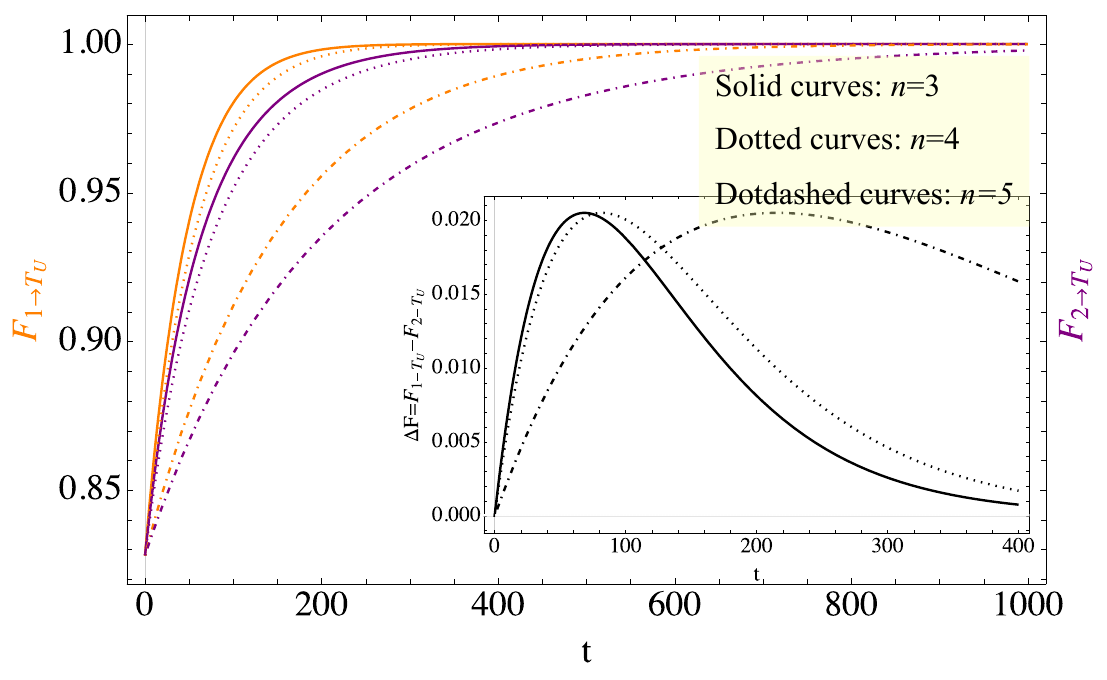}}\\
            \subfloat[Massless scalar]{\includegraphics[width=.48\textwidth]{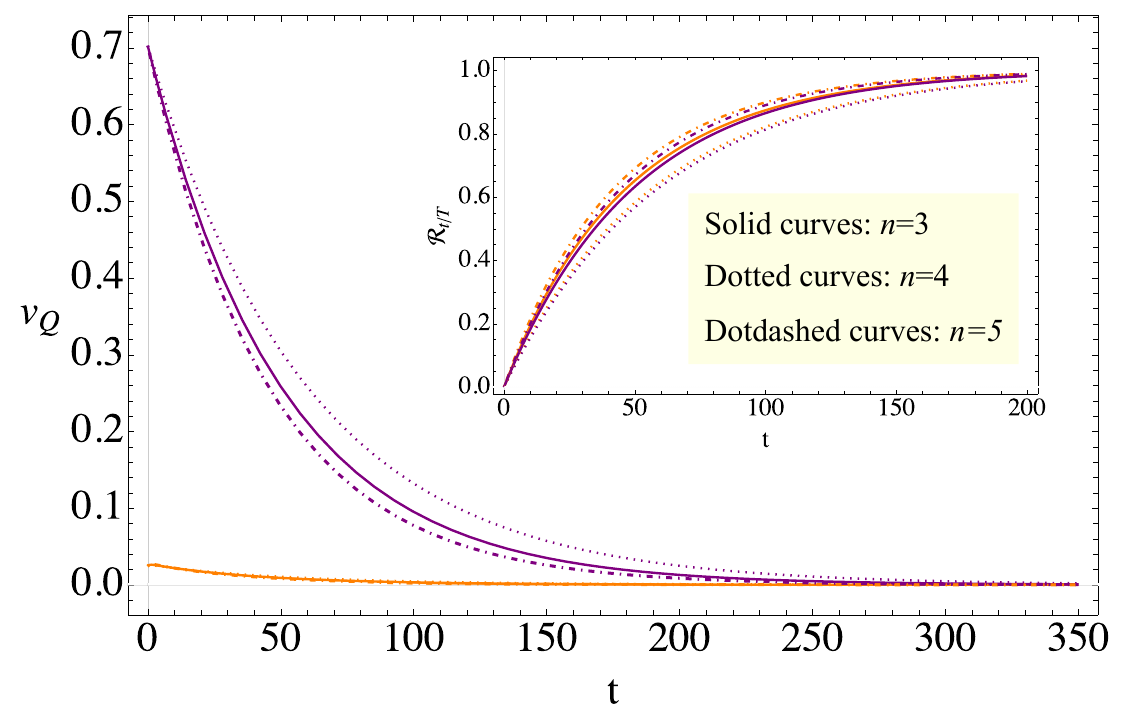}}
        \subfloat[Thermal field]{\includegraphics[width=.47\textwidth]{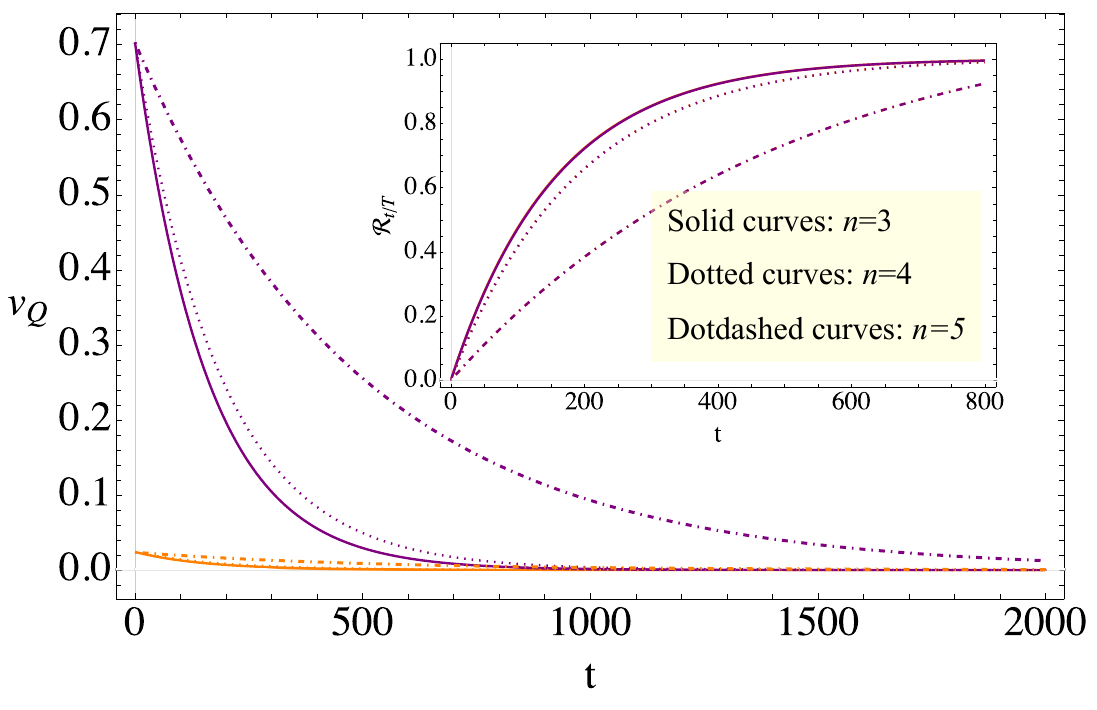}}
    \caption{The time-evolution of fidelity, velocity, and QDC for two UDW detectors starting from non-thermal states with $(l_0^{(1)},\Theta_0^{(1)})=(0.0258,2)$ and $(l_0^{(2)},\Theta_0^{(2)})=(0.773,2)$, both has equal fidelity $F=0.782$ to the thermalization end characterized by Unruh temperature $T_U=0.5$.}
    \label{G1-fidelity-velocity-QDC}
\end{figure}

\begin{figure}[htbp]
\centering
    \subfloat[Massless scalar]{\includegraphics[width=.48\textwidth]{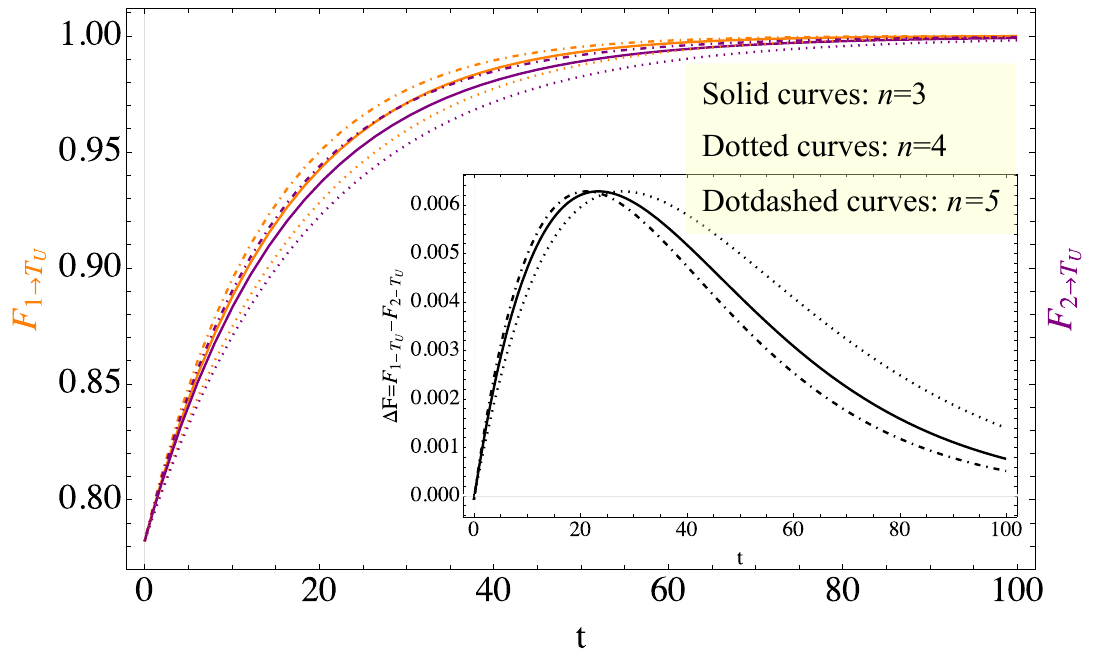}}
        \subfloat[Thermal field]{\includegraphics[width=.48\textwidth]{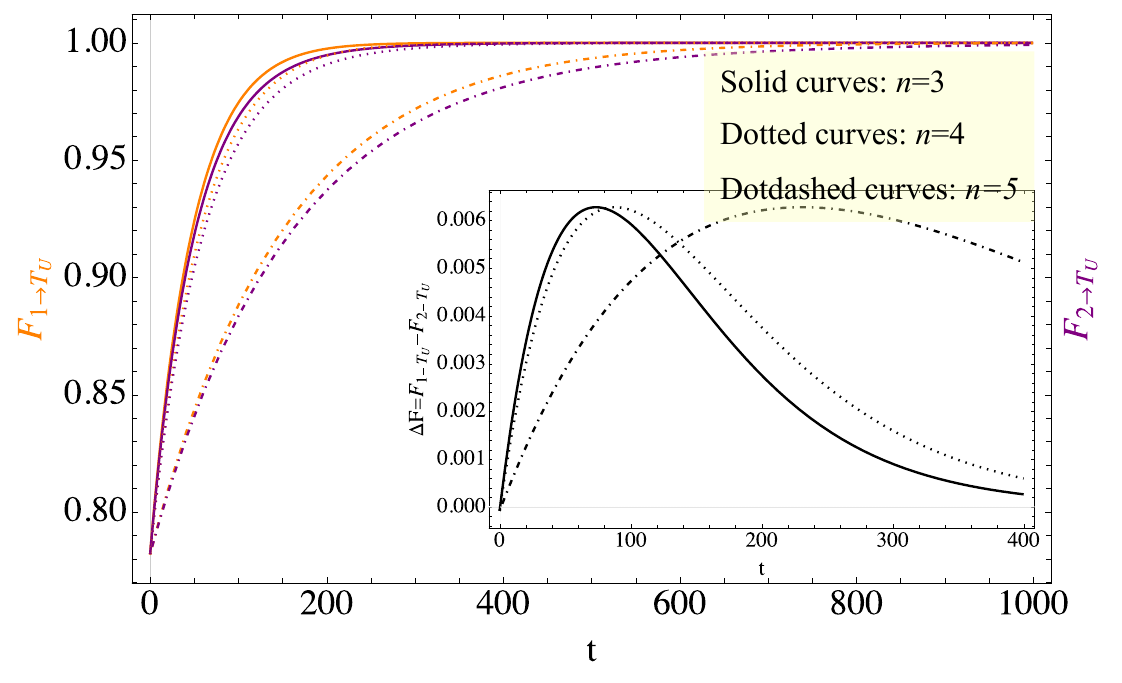}}\\
            \subfloat[Massless scalar]{\includegraphics[width=.47\textwidth]{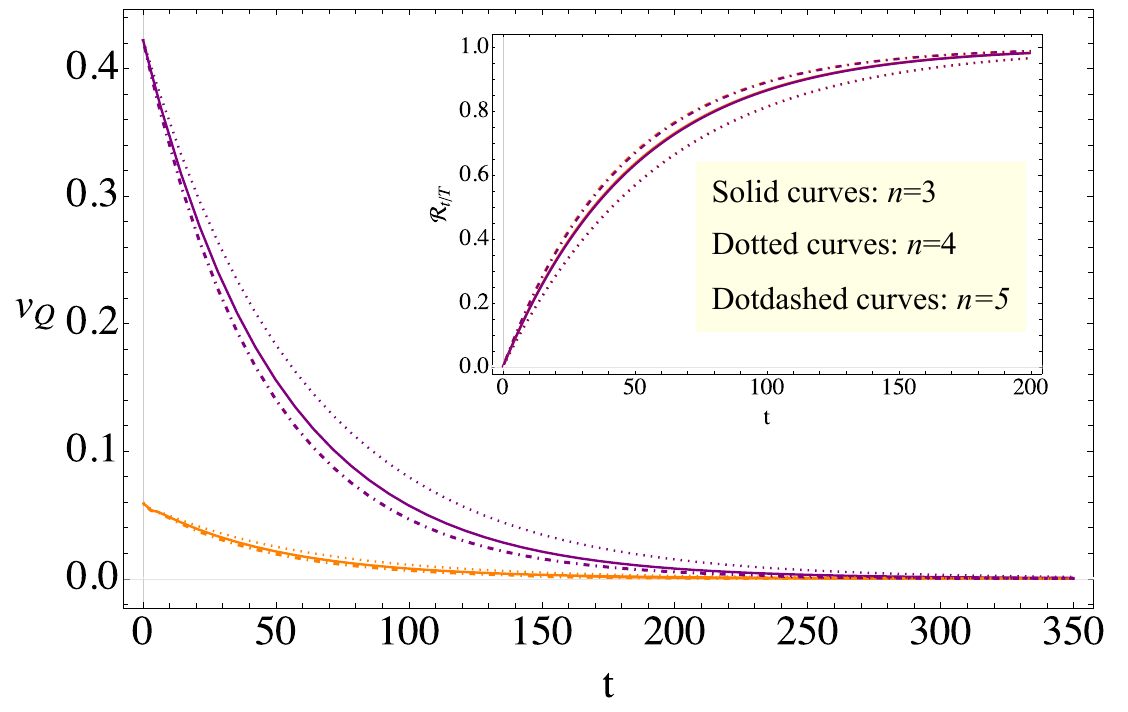}}~
        \subfloat[Thermal field]{\includegraphics[width=.47\textwidth]{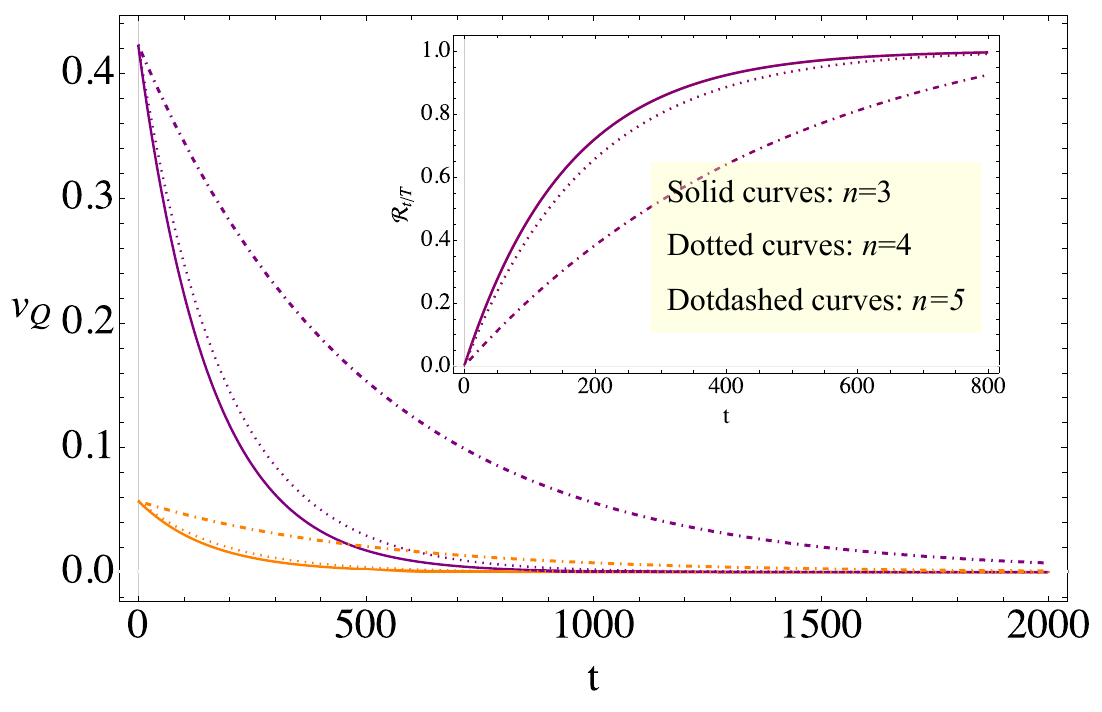}}
    \caption{The time-evolution of fidelity, velocity, and QDC for two UDW detectors starting from initial non-thermal states with $(l_0^{(1)},\Theta_0^{(1)})=(0.119,0.5)$ and $(l_0^{(2)},\Theta_0^{(2)})=(0.424,1.5)$, both has equal fidelity $F=0.828$ to the thermalization end characterized by Unruh temperature $T_U=0.5$}
    \label{G2-fidelity-velocity-QDC}
\end{figure}

First, we consider two UDW detectors that are initially prepared with $(l_0^{(1)},\Theta_0^{(1)})=(0.0258,2)$ and $(l_0^{(2)},\Theta_0^{(2)})=(0.773,2)$, both has equal fidelity $F=0.782$ to the thermalization end characterized by Unruh temperature $T_U=0.5$. We depict in Fig.\ref{G1-fidelity-velocity-QDC} the time evolution of the fidelity, quantum velocity, and the QDC of detector thermalization processes $\rho_i^{(1)}\rar\rho_{\text{eq}}$ (orange curves) and $\rho_i^{(2)}\rar\rho_{\text{eq}}$ (purple curves) within different baths background. Generally, we observe that the fidelity of one process consistently exceeds that of another, similar to the previously discussed three-temperature protocol, but without any concepts of "heating" or "cooling" that can be defined for the process starting from nonthermal states. Compared to the Unruh thermalization within a scalar quantum field (Fig.\ref{G1-fidelity-velocity-QDC}(a)(c)), we find significantly longer timescales for the thermalization process within a classical bath (Fig.\ref{G1-fidelity-velocity-QDC}(b)(d)).

Second example is the two UDW detectors that are initially prepared with $(l_0^{(1)},\Theta_0^{(1)})=(0.119,0.5)$ and $(l_0^{(2)},\Theta_0^{(2)})=(0.424,1.5)$, both has equal fidelity $F=0.828$ to the thermalization end characterized by Unruh temperature $T_U=0.5$. We depict in Fig.\ref{G2-fidelity-velocity-QDC} the time evolution of the fidelity, velocity, and the QDC of detector thermalization processes $\rho_i^{(1)}\rar\rho_{\text{eq}}$ (orange curves) and $\rho_i^{(2)}\rar\rho_{\text{eq}}$ (purple curves) within different baths background. We see that the time evolution of fidelity, velocity, and QDC presented in Fig.\ref{G2-fidelity-velocity-QDC} exhibits no substantial differences from that demonstrated in Fig.\ref{G1-fidelity-velocity-QDC}.

It is noteworthy that the difference in velocity or QDC caused by increasing spacetime dimensions (Fig.\ref{G1-fidelity-velocity-QDC}(c)(d) and Fig.\ref{G2-fidelity-velocity-QDC}(c)(d)) is considerably smaller compared to the three-temperature protocol. This is because, for detectors starting with non-thermal states, the initial condition $\Theta_0\neq n\pi$ causes the QFI \eqref{QFI2t} to include a nonvanishing coherent contribution $\mathcal{I}_Q^{coh}$, which greatly suppresses the variation of QFI for different spacetime dimensionality.

\section{Conclusions}
\label{sec5}

In this paper, we analyze the complete open quantum dynamics of a UDW detector interacting with typical background fields in the $n$-dimensional Minkowski spacetime. From a quantum thermodynamics perspective, we describe the irreversible thermalization process of the detector on the Bloch sphere, as it undergoes Unruh or thermal radiation. We observe several signatures that distinguish Unruh thermalization from the process driven by a thermal bath. First, we find that the timescale of thermalization with a classical bath is significantly longer than the scale of the Unruh thermalization. Second, the difference between the change rates of quantum coherence and heat $\Delta(t ; n):=\dot{\mathbb{Q}}-\dot{\mathbb{C}}$ can serve as a compelling diagnosis of Unruh thermalization from its classical counterpart, by examining its maximal value for different spacetime dimensions. We also investigate the quantum thermal kinematics of the heating/cooling protocols for the UDW detector, utilizing tools from information geometry. Our main finding is that, while a quantum Mpemba-like effect (QME) of Unruh thermalization is observed, the maximum of fidelity difference $\Delta F:=F_{\text {heating }}-F_{\text {cooling }}$ provides another compelling signature that essentially distinguishes the thermalization process induced by the Unruh effect from that caused by a classical thermal bath.   

As aforementioned, a key experimental evidence of the genuine Unruh effect is a quantum-discriminating signature. With recent progress in simulating the UDW detector undergoing Unruh radiation \cite{v2-12,sec5-0} and the quantum Mpemba effect in non-equilibrium many-body systems \cite{sec5-1,sec5-2,sec5-3}, we expect that the $\Delta(t ; n)$ and $\Delta F$ can serve as a new criterion for future experimental tests of the Unruh effect.

We want to discuss several possible extensions of our work. First, while we assume the detectors' dynamics are isolated in the three-temperature protocol, it could be interesting to incorporate mutual influence via the common field they are coupled to. This would require treating the joint $4\times4$ dynamics (or an appropriate correlated master equation), and possible \emph{entanglement harvesting} between the detectors \cite{rev-4} may play a non-negligible role. Second, beyond linear acceleration, the detailed balance temperatures along other stationary, uniformly accelerated trajectories exhibit a complicated dependence on additional geometric parameters (such as torsion) \cite{rev-6}. From an experimental viewpoint, typical circular motion allows the accelerating system to remain within a finite-size laboratory for an arbitrarily long interaction time, making it an attractive candidate for measuring the Unruh effect \cite{rev-7}. Extending our work to such a generalized Unruh effect, especially to identify related quantum-discriminating signatures, as well as the Mpemba-like effect, may have significant experimental relevance. Finally, in a more realistic setting, the sudden quench we assumed in two/three-temperature protocols may fail, as the detector should be taken along a particular (nonstationary) trajectory that actually interpolates between the two relevant temperatures. A systematic treatment of Mpemba-type behavior under such nonstationary ramps is surely important for realistic reasons. However, potentially non-Markovian open dynamics \cite{rev-8} and additional control parameters characterizing trajectory geometry, as can be foreseen, make such attempts highly challenging.


\section*{Acknowledgements}
This work is supported by the National Natural Science Foundation of China (No. 12475061) and the Shaanxi Fundamental Science Research Project for Mathematics and Physics (No. 23JSY006), and the Innovation Program for Quantum Science and Technology (2021ZD0302400). J.F. would especially like to thank Yeting Wen and Zi-Yu Feng for their support throughout the completion of this work.

\end{document}